\documentclass[12pt, letterpaper]{article}

\usepackage{amsmath,amssymb}
\usepackage{cite}
\usepackage{fancyhdr}
\usepackage[top=1in, bottom=1in, left=1in, right=1in]{geometry}
\usepackage{graphicx}
\usepackage{hyperref}
\usepackage{sidecap}

\numberwithin{equation}{section}

\begin{document}


\setcounter{page}{0}
\date{}

\lhead{}\chead{}\rhead{\footnotesize{RUNHETC-2013-11\\SCIPP-13/07\\UTTG-13-13\\TCC-012-13}}\lfoot{}\cfoot{}\rfoot{}

\title{\textbf{Holographic Fluctuations from Unitary de Sitter Invariant Field Theory\vspace{0.4cm}}}

\author{Tom Banks$^{1,2}$ \and Willy Fischler$^{3}$ \and T.J. Torres$^{2}$ \and Carroll L. Wainwright$^{2}$
\vspace{0.7cm}\\
{\normalsize{$^1$NHETC and Department of Physics and Astronomy, Rutgers University}}\\
{\normalsize{Piscataway, NJ 08854-8019, USA}}\vspace{0.2cm}\\
{\normalsize{$^2$SCIPP and Department of Physics, University of California,}}\\
{\normalsize{Santa Cruz, CA 95064-1077, USA}}\vspace{0.2cm}\\
{\normalsize{$^3$Department of Physics and Texas Cosmology Center}}\\
{\normalsize{University of Texas, Austin, TX 78712}}\\
{\normalsize{E-mail: \href{mailto:fischler@physics.utexas.edu}{fischler@physics.utexas.edu}}}}

\maketitle
\thispagestyle{fancy} 

\begin{abstract}
\normalsize \noindent
We continue the study of inflationary fluctuations in Holographic Space Time models of inflation. We argue that the holographic theory of inflation\cite{holoinflation} provides a physical context for what is often called dS/CFT.  The holographic theory is a quantum theory which, in the limit of a large number of e-foldings, gives rise to a field theory on $S^3$, which is the representation space for a unitary representation of $SO(1,4)$.  This is {\it not} a conventional CFT, and we do not know the detailed non-perturbative axioms for correlation functions.  However, the two- and three-point functions are completely determined by symmetry, and coincide up to a few constants (really functions of the background FRW geometry) with those calculated in a single field slow-roll inflation model.  The only significant deviation from slow roll is in the tensor fluctuations. We predict zero tensor tilt and roughly equal weight for all three conformally invariant tensor 3-point functions (unless parity is imposed as a symmetry). We discuss the relation between our results and those of \cite{malda}, \cite{mcfaddenskenderis} and \cite{others}.
Current data can be explained in terms of symmetries and a few general principles, and is consistent with a large class of models, including HST. \end{abstract}


\newpage
\tableofcontents
\vspace{1cm}

\section{Introduction}

Holographic Space Time (HST) is a formalism for generalizing string theory to situations where the asymptotic regions of space-time are not frozen vacuum states. 
In particular, it gives us a well defined holographic quantum theory of Big Bang cosmology\cite{holocosm}.  In \cite{holoinflation}, two of the authors (TB and WF) introduced a model of HST, which they claimed could reproduce the results of slow-roll inflation.  In this paper, we use results from \cite{malda},\cite{mcfaddenskenderis} and \cite{others} to prove and improve that claim.  As a consequence we will show that {\it if the two-point functions of inflationary fluctuations coincide with those in a single-field slow-roll model, then they probe only coarse features of the underlying fundamental quantum theory. }  Any model which produces small, approximately Gaussian, approximately $SO(1,4)$ covariant fluctuations yields two- and three-point functions determined by two unitary representations of $SO(1,4)$.  A generic model has $9$ parameters: the scaling dimension of the scalar operator on the projective light cone (see below); the strength of the scalar and tensor Gaussian fluctuations; the normalizations of the $\langle S^3\rangle$, $\langle ST^2\rangle$, and $\langle S^2 T\rangle$ three-point functions; and the 3 different tensor structures in the $\langle T^3 \rangle$ three-point function.   Maldacena's squeezed limit theorem, when combined with $SO(1,4)$, fixes all 3-point functions involving the scalar in terms of the scale dependence of the corresponding two-point functions, reducing the number of parameters to $6$\footnote{We emphasize that the word parameters above actually refers to functions of the background Hubble radius $H(t)$ and its first two time derivatives.}.  A general quantum theory with $SO(1,4)$ invariance and localized operators $S$ and $T$ contains many invariant density matrices, and the two- and three-point functions do not determine either the underlying model or the particular invariant density matrix.  Finally, we note that the dominance of the scalar over the tensor two-point fluctuations, which is evident in the CMB data, follows from general cosmological perturbation theory along with the assumptions that there was a period of near dS evolution and that the intrinsic $S$ and $T$ fluctuations are of the same order of magnitude, which is the case in both conventional slow-roll models and the HST model of inflation.  However, HST predicts a different dependence of the fluctuation normalizations on the time dependent background $H(t)$.  For correlation functions involving the scalar, this difference can be masked by choosing different backgrounds to fit the data.  HST makes the unambiguous prediction that the tensor tilt vanishes, whereas conventional slow roll predicts it to be $r/8$ (where $r$ is the ratio of amplitudes of tensor and scalar fluctuations) and data already constrains $r < 0.1$, so this may be hard to differentiate from zero even if we observe the tensor fluctuations.

We have noted a set of relations for fluctuations that follow from very general properties of cosmological perturbation theory.  For us, the validity of classical cosmological perturbation theory follows from Jacobson's observation that Einstein's equations (up to the cosmological constant, which in HST is a boundary condition for large proper times) are the hydrodynamic equations of a system, like HST, which obeys the Bekenstein-Hawking relation between entropy and area. In \cite{holoinflation} we argued that this meant that there would be a hydrodynamic inflaton field (or fields) even in regimes where HST is not well-approximated by {\it quantum} effective field theory (QUEFT)\footnote{We refer to the classical field equations, which, according to Jacobson, encode the hydrodynamics of space-time as a Thermodynamic Effective Field Theory, or THEFT.}.  The fluctuations calculated from an underlying HST model, which we argued to be approximately $SO(1,4)$ covariant and whose magnitude we estimated, are put into the classical hydrodynamical inflaton equations as fluctuating initial conditions.  In fact, in the co-moving gauge, we can view the inflaton as part of the metric and this picture follows from Jacobson's original argument\footnote{Even in the FRW part of the metric, we can view the inflaton field as just a generic way to impose the dominant energy condition on a geometry defined by an otherwise unconstrained Hubble radius $H(t)$.}.  

We will argue below that certain constraints on the parameters not determined by symmetries follow from quite general arguments based on classical cosmological perturbation theory, while other constraints correspond to a choice of parameters in the underlying discrete HST model. Note that, within the HST framework, the validity of classical cosmological perturbation theory is a statement about the Jacobsonian hydrodynamics of a system, which is NOT well-approximated by QUEFT.  The constraints fix the $SO(1,4)$ representation of the tensor fluctuations, and imply the dominance of scalar over tensor fluctuations. When combined with estimates from the HST model, these constraints suggest that the tensor two-point function {\it  might} be observed in the Planck data.  Non-gaussian fluctuations involving at least one scalar component are small.  The most powerful discriminant between models is the tensor 3 point function.  Standard slow-roll models produce only one of the three forms allowed by symmetry.  A second one can be incorporated by adding higher derivative terms to the bulk effective action, but the validity of the bulk effective field theory expansion requires it to be smaller than the dominant term\footnote{In weakly coupled string theory, we might consider inflation with Hubble radius at the string scale, and get this second term to be sizable, but this is a non-generic situation, based on a hypothetical weakly coupled inflation model, for which we do not have a worldsheet description.}.  The third form violates parity and cannot arise in any model based on bulk effective field theory.  We argue that it could be present in more general models, if parity is not imposed as a fundamental symmetry.
Unfortunately, all extant models, whether based on effective field theory or HST, predict that the tensor 3 point function is smaller than the, as yet unobserved, tensor two-point function by a power of $\frac{H}{m_P}$ and we cannot expect to detect it in the foreseeable future.

This paper is organized as follows:  in the next section we present the mathematical analysis of two- and three-point functions in a generic quantum theory carrying a unitary representation of $SO(1,4)$, written in terms of operators localized on a three sphere.
We emphasize that this theory does not satisfy the axioms of quantum field theory on the three sphere. In particular, it does not satisfy reflection positivity, because the Hamiltonian is a generator of a unitary representation of $SO(1,4)$, and it is not bounded from below. The absence of highest weight generators prevents the usual continuation to a Lorentzian signature space-time.  In consequence, a general theory of this type will have a large selection of dS invariant density matrices, rather than the unique pure state of conventional CFT. Nonetheless, two- and three-point functions are completely determined by symmetry, up to a few constants.  The work of Maldacena and collaborators\cite{malda} (see also later work of Mcfadden and Skenderis\cite{mcfaddenskenderis} and others\cite{others}) shows that, to leading order in slow-roll parameters, single field slow-roll inflation is in this category of theories\footnote{When we use the phrase slow-roll model, we mean a model in which the fluctuations are calculated in terms of QUEFT in a slow-roll background. In our HST model, the fluctuations are calculated in an underlying non-field theoretical quantum model, and put into the hydrodynamic equations of that model as initial conditions.}.  Thus, the oft-heard claim that objections\footnote{We recall these objections in  Appendix A.} to the conceptual basis of slow-roll inflation must be wrong, because the theory fits the data, are ill-founded.  Our analysis shows that current data probe only certain approximate symmetries of a theory of primordial fluctuations, and determine a small number of parameters that are undetermined by group theory.  Furthermore, the success of the slow-roll fit to these parameters, amounts, so far, to the statement that the fluctuations are predicted to be small and approximately Gaussian; that the scaling exponents are within a small range around certain critical values; and that the scalar two-point function is much larger than that of the tensor.  Given the central limit theorem, the first part of this prediction does not seem to be such an impressive statement.  Note also that if there is any environmental selection going on in the explanation  of the initial conditions of the universe, even the statement that the fluctuations are small might be understood as environmental selection.  The fact that the scalar two-point function dominates the tensor is a consequence of general properties of classical gravitational perturbation theory around a background which is approximately de Sitter.
The relative sizes of various three-point functions were also derived by Maldacena in this quite general setting.  

We should emphasize that our remarks are relevant to data analysis only if future data remains consistent with slow-roll inflation.  There is a variety of inflationary models, and many of them produce non-Gaussian fluctuations which are not $SO(1,4)$ covariant.  If future observations favor such a model, they would rule out the simple symmetry arguments, and disprove both slow-roll inflation and the holographic model of inflation. Thus, although we set out to prove that our HST model {\it could} fit the data, we have ended up realizing that the current data probe only a few general properties of the underlying theory of primordial fluctuations.  At the moment, we do not have enough control over our model to make predictions that go beyond these simple ones.  

In section 3 we sketch the holographic inflation (HI) model of \cite{holoinflation} and recall how it leads to a prediction of small, approximately Gaussian, approximately $SO(1,4)$ invariant fluctuations.  It also resolves all of the conceptual problems of conventional inflation and gives a completely quantum mechanical and causal solution to the flatness and horizon problems, as well as an explanation of the homogeneity and isotropy of the very early universe --- all of the latter without inflation.  Within the HST formalism, the HI model also explains why there is any local physics in the world, despite the strong entropic pressure to fill the universe with a single black hole at all times.

In the Conclusions we discuss ways in which the data might distinguish between different models of fluctuations.  Appendix A recalls the conceptual problems of conventional inflation models, while Appendix B recalls the unitary irreducible representations of $SO(1,4)$.

\section{Fluctuations from symmetry}

In early work on holographic cosmology\cite{holocosm} TB and WF postulated that inflation took place at a time when the universe was well described by effective quantum field theory (QUEFT), and that the inflaton was a quantum field.  Our attitude to this began to change as a consequence of two considerations.  The first was that, although inflationary cosmology and de Sitter space are {\it not} the same thing, it seemed plausible that at least part of the fully quantum version of inflationary cosmology should involve evolution of independent dS horizon volumes in a manner identical to a stable dS space, over many e-foldings.  However, over such time scales we expected each horizon volume to be fully thermalized.  The black hole entropy formulas in dS space tell us that the fully thermalized state has {\it no local excitations} and therefore is not well modeled by field theory.

In parallel with this realization, we began to appreciate Jacobson's 1995 argument\cite{ted}, indicating that the classical Einstein equations were just hydrodynamics for a system obeying the local connection between area and entropy, for maximally accelerated Rindler observers.  The gravitational field should only be quantized in special circumstances where the covariant entropy bound is far from saturated and bulk localized excitations are decoupled from most of the horizon degrees of freedom (DOF). Jacobson's argument does not give a closed system of equations, because it does not provide a model for the stress tensor.  We realized that this meant that other fields like the inflaton, which provide the stress tensor model, could also be classical hydrodynamical fields, unrelated to the QUEFT fields that describe particle physics in the later stages of the universe.

In \cite{holoinflation} TB and WF constructed a model which begins with a maximal entropy $p=\rho$ Big Bang, passes through a stage with $e^{3N_e}$ decoupled horizon volumes of dS space and evolves to a model with approximate $SO(1,4)$ invariance.  The corrections to $SO(1,4)$ for correlation functions of a small number of operators are of order $e^{-N_e}$. In trying to assess the extent to which the predictions of such a model could fit CMB and large scale structure data, we realized that, to leading order in the slow-roll approximation, {\it the results of many conventional inflationary models amounted to a prediction of approximate $SO(1,4)$ invariance and the choice of a small number of parameters.}  Work of \cite{malda},\cite{mcfaddenskenderis} and \cite{others} has shown that even if we can measure all two- and three-point functions of both scalar and tensor fluctuations, there are only $9$ parameters.  Some of these parameters are related, by an argument due to Maldacena.  Current measurements only determine two of the parameters and bound some of the others
(the tensor spectral index can't be measured until we actually see tensor fluctuations).  Our conclusion is that observations and general principles tell us only that the correct theory of the inflationary universe has the following properties

\begin{itemize}

\item It is a quantum theory that is approximately $SO(1,4)$ invariant, and the density matrix of the universe at the end of inflation is approximately invariant.  There are many such density matrices in a typical reducible representation of $SO(1,4)$.

\item The tensor and scalar fluctuations are expectation values of operators transforming in two particular representations of $SO(1,4)$.  CMB and LSS data determine the normalization of the two-point function and representation
of the scalar operator, and put bounds on the two-point function of the tensor and all 3 point functions.  If future measurements detect neither B mode polarization nor indications that the fluctuations are non-Gaussian, then we will learn no more about the correct description of the universe before and during the inflationary era.

\item When combined with Maldacena's ``squeezed limit" theorem and general features of cosmological perturbation theory around an approximately dS solution, $SO(1,4)$ invariance gives results almost equivalent to a single field slow-roll model.  We will discuss the differences below.  Thus, it is possible that even measurement of {\it all} the two- and three-point functions will teach us only about the symmetry properties of the underlying model. In fact, measurement of the tensor 3-point function could rule out conventional slow roll, but would not distinguish between more general models obeying the symmetry criteria described above.

\end{itemize}

In particular, the models proposed in \cite{holoinflation}, which resolve all of the conceptual problems of QUEFT based inflation models (see Appendix A), 
will fit the current data as well as any conventional model.  At our current level of understanding, those models do not permit us to give a detailed prediction for the scalar tilt, apart from the fact that it should be small.  The tensor tilt is predicted to vanish.  HST models suggest very small non-Gaussianity in correlation functions involving scalar perturbations, as we will see below, and explain why the scalar two-point function is much larger than that of the tensor.  Indeed, this follows from $SO(1,4)$ symmetry, Maldacena's long wavelength theorem for scalar fluctuations, and very general properties of classical gravitational fluctuations around a nearly dS FRW model.

Classical cosmological perturbation theory identifies two gauge invariant quantities, which characterize fluctuations, and transform as a scalar and a transverse traceless tensor under $SO(3)$.  We will attempt to find $SO(1,4)$ covariant operators, whose expectation values give us the two- and three-point correlation functions of these fluctuations.  The form of these fluctuations is determined by group theory, up to a few constants.

In order to handle $SO(1,4)$ transformation properties in a compact manner, we use the description of the 3-sphere as the projective future light cone in $4 + 1$ dimensional Minkowski space.  That is, it is the set of $5$ component vectors $X^{\mu}$ satisfying
$X^{\mu} X_{\mu} = 0$, $X^0 > 0$ and identified under $X^{\mu} \rightarrow \lambda X^{\mu}$ with $\lambda > 0$.  Fields on the sphere are $SO(1,4)$ tensor functions of $X$, which transform as representations of the group of identifications, isomorphic to $R^+$.  These representations are characterized by their tensor transformation properties, covariant constraints, and a single complex number $\Delta$
such that $$ F (\lambda X) = \lambda^{- \Delta} F(X), $$ so that the field is completely determined by its values on the sphere.  The allowed values of $\Delta$ are constrained by the unitarity of the representation of $SO(1,4)$ in the Hilbert space of the theory.  An expectation value of the products of two or three of these operators, in any dS invariant density matrix, is determined in terms of the $9$ numbers discussed above.  For the tensor modes, the determination of the 3 point function has been demonstrated in \cite{malda} and \cite{others}, but there is not yet a compact formula.  We hope that the five dimensional formalism will provide one, but we reserve this for future work.

Ordinary QFT in dS space has {\it many} dS invariant states, both pure and impure, as a consequence of the fact that there are no highest weight unitary representations of $SO(1,4)$\footnote{We are not talking here about the (in)-famous $\alpha$ vacua, which are states of Gaussian quantum fields whose two-point function is dS invariant, but has singularities when a point approaches its anti-pode.  Rather, in the context of bulk QUEFT, we're speaking about dS invariant excitations of the conventional Bunch-Davies vacuum. These are not represented by Gaussian wave functionals.}. In conventional CFT, the product of two lowest weight representations contains only representations of weight higher than the sum of the individual weights, but this is not true for unitary representations of $SO(1,4)$. Products of non-trivial irreps can have singlet components.  We can make even more general $SO(1,4)$ invariant density matrices by taking weighted sums of the projectors on this plethora of pure invariant states.  This is in marked contrast to conventional CFT, whose Hilbert space consists of lowest weight unitary representations of $SO(2,3)$ and has a unique invariant state.  Nonetheless, because the constraints of dS invariance on $2$ and $3$ point functions are expressed as analytic partial differential equations, in which the cosmological constant appears as an analytic parameter, these functions {\it are} analytic continuations of corresponding expressions in ordinary CFT.  While we do not think that QFT in dS space is the correct theory of inflationary fluctuations, nor that the quantum theory of dS space is $SO(1,4)$ invariant; we do think that it is plausible that the quantum theory of a cosmology that has a large number of e-folds of inflation, followed by sub-luminal expansion which allows observers to see all of that space-time, should have an approximate $SO(1,4)$ symmetry realized by unitary operators in a Hilbert space.  This was explained in \cite{holoinflation}.

The scaling symmetry $R^+$ plays another useful role, since we are trying to make a quantum model of many horizon volumes of the asymptotic future of a classical dS space.  dS space asymptotes to the future light cone $X^2 = 0$, and the rescaling transformation is simply time evolution in either the global or flat slicing.  The two times are asymptotically equivalent. 
For large time in the flat slicing, we have $$X\cdot Y \sim e^{2t/R} ({\bf x - y})^2 .$$  Thus, the scaling dimension of the operator tells us about its large time behavior in the flat slicing of dS space.

  The field corresponding to the scalar fluctuations is a scalar $S (X)$, with $\Delta_S = (3/2 - \sqrt{3/2 - m^2 R^2})$.  In this formula $m^2$ is the mass of a bulk scalar field, which would give rise to this representation by dS/CFT calculations, as in \cite{malda}, when evaluated in the Bunch-Davies vacuum.  It parametrizes one of the series (Called Class I in Appendix B) of unitary representations of $SO(1,4)$ described in \cite{thomasnewton}. These are the analogs of the complementary and principal series of unitary representations of $SL(2,R)$.  In ordinary QFT in dS space, the Wheeler-DeWitt wave function for this bulk field determines the in-in correlator of the corresponding bulk quantum field in the Bunch-Davies vacuum. We note that this is the result of direct calculation of ordinary QFT in dS space and does not invoke any analytic continuation from an AdS calculation.  We will discuss the relation to dS/CFT in more detail below.  The values for which the square root is real are called the complementary series of representations, while those for which the real part is fixed at $3/2$ while the imaginary part varies, are the principal series.  At the level of the two-point function we could view the usual scalar fluctuation as determined by the correlator of $\langle \Phi^{\dagger} \Phi \rangle$, which makes sense even for the complex operators of the principle series.  However, there is no consistent interpretation of the 3 point function of principal series operators, except to set the various complex 3 point functions to zero.  Thus, in order to have an interpretation as fluctuations of the real quantity $\zeta$, we must restrict attention to the complementary series.  Thus $\Delta_S$ is bounded between $0$ and $3/2$. Conventional slow-roll models have $\Delta_S = 0$.  Note that, strictly speaking, this is not in the list of unitary representations, but is a limit of them.  We believe that this is a consequence of the logarithmic behavior of massless minimally coupled propagators in dS space, which may make the definition of the global symmetry generators a bit delicate.
  From the point of view of phenomenology, this subtlety is irrelevant.  $SO(1,4)$ invariance is only an approximate property of inflationary models and we can certainly consider arbitrarily small values of $\Delta_S$, so the distinction between zero and other values could never be determined from the data.

Our point is that the representation constant of the field fixes its two and 3 point functions up to pre-factors. The two-point function of such a field is fixed, up to a multiplicative constant to be
$$ {\rm Tr} [\rho \Phi (X) \Phi (Y) ] = C^S_2 (X^{\mu} Y_{\mu} )^{- 2 \Delta_S} .$$ Here $\rho$ can be any $SO(1,4)$ invariant density matrix.

The flat slicing of dS space is
$$ ds^2 =  - dt^2 + e^{2t/R}\ d {\bf y}^2  .$$ Using the asymptotic relation between the flat coordinates and the light cone, we find a momentum space correlator (in radial momentum space coordinates)
$$ 4\pi C^S_2 k^{-1}( \frac{e^{t/R}}{k})^{ 4\Delta_S} .$$

Similarly, the 3 point function is determined to be
$${\rm Tr} [ \rho \Phi (X_1 , X_2 , X_3 ) ]  = C_3 \prod_{ij} X_{ij}^{\frac{-\Delta_S}{2}} ,$$ where
$X_{ij} = X_i \cdot X_j .$  This form follows from $SO(1,4)$, the scaling symmetry, and symmetry under permutations of the points. The latter symmetry follows from the assumption that the operators commute with each other.  In both HST and a conventional slow-roll model, this is a consequence of the fact that the different points are causally disconnected during inflation, and that we are computing an expectation value at fixed time.  In the HST model, $SO(1,4)$ invariance sets in via a coupling together of DOF at different points, using a time dependent Hamiltonian, which approaches a generator of $SO(1,4)$ when the number of e-folds is large\footnote{Below, we'll recall the meaning of the bulk concept ``number of e-foldings" in the HST model.}.

Maldacena and Pimentel\cite{malda}, among others \cite{others}, have shown how all of the $3$ point functions are determined up to 3 normalizations by $SO(1,4)$ group theory.  Similar results were obtained by McFadden, Skenderis and collaborators\cite{mcfaddenskenderis}.  Our results for correlation functions of tensor modes have to coincide with theirs because the only representation of $SO(1,4)$ that has the right number of components to represent the transverse traceless graviton fluctuation is the Class IV representation (in the notation of Newton\cite{thomasnewton}) with $s=2$. The Casimir operators have the values  $Q = -6 , W = -12$.  The class $IV_{a,b}$ representations are the two different helicity modes of the graviton.  See Appendix B for details of the classification of $SO(1,4)$ representations.  Note however that the coefficients in front of these group theoretic predictions are different in slow-roll and HST models.  In slow-roll models, both scalar and tensor fluctuations are computed as two-point functions of quantum fields in the background space-time $H(t)$, while in HST, the magnitude of the fluctuations is determined by a fixed Hubble parameter $H$, as we will review below.  

Two interesting points about the pure tensor 3-point functions were made by \cite{parityviolating} and by Maldacena and Pimentel\footnote{The explicit forms for these three-point functions are not terribly illuminating.
The most elegant expression we know is in the spinor helicity formalism used by Maldacena and Pimentel, and it would be redundant and pointless to reproduce that here.  We hope that the realization of the three sphere as the five dimensional projective light cone will simplify these expressions, but we have not yet succeeded in showing this.}. In bulk field theory computations, the parity violating term allowed by group theory does not appear in correlation functions.  The corresponding term in the logarithm of the WD wave function is purely imaginary and does not contribute to correlators of operators that are simply functionals of the fields appearing in the wave functional.  Neither the tensor nor scalar operators involve functional derivatives acting on the wave functional, and so their correlators are insensitive to this term. In addition, one of the two parity conserving structures only appears if we allow higher derivative terms in the bulk action.  In a more general $SO(1,4)$ invariant theory, the vanishing of the parity violating term might follow from an underlying reflection invariance of the microscopic dynamics, while there is no reason for the two parity conserving terms to have very different normalizations\footnote{Maldacena and Pimentel point out that in a hypothetical model of inflation in perturbative string theory, the derivative expansion can break down even though quantum gravity corrections are negligible, if the Hubble scale is the string scale.  They argue that this could produce parity conserving terms, with comparable magnitude. An actual computation of these terms would
require us to find a worldsheet formulation of the hypothetical weakly coupled string model of inflation.}.  

In general, knowledge of the parity operation on the fields, plus the fact that the fields commute with each other, {\it does not} imply that the parity operator commutes with the fields.  Rather, it is like a permutation operator, which permutes the elements of a complete ortho-normal basis.  The properties of parity imply that it squares to a multiple of the unit operator.
In the conventional approach to slow-roll inflation models, the Hilbert space is interpreted as the thermo-field double of field theory in a single causal patch and the state is taken to be the Bunch-Davies state, which reproduces thermal correlation functions in the theory of the causal patch.  This state is invariant under a $Z_2$ which reverses both the orientation of the 3-sphere and the time in the causal patch.  When combined with the TCP invariance of bulk quantum field theory, this leads to parity invariant correlation functions.  Another way of seeing the same result is to note that the WD density matrix for the Bunch-Davies vacuum, is diagonal in the same basis as the fields whose expectation values we are computing.  The parity operation is defined as complex conjugation of the WD wave function, and leaves the diagonal matrix elements of the density matrix invariant.  These are the only matrix elements relevant to calculating these particular expectation values.

In the HI model of inflation, thermal fluctuations in {\it many} initially decoupled dS causal patches are coupled together by a time dependent Hamiltonian, which, in the limit of a large number of e-folds, approaches a generator of $SO(1,4)$.  In this limit, one can argue that the density matrix should become approximately $SO(1,4)$ invariant, but we do not see a general argument that it be parity invariant. Similarly, there is no reason for the density matrix to be diagonal in the same basis as the fields $S(X)$ and $T(X)$ on the three sphere.  The parity operation acts simply on the fields, but not necessarily on the density matrix. {\it Consequently, there is no argument that the parity violating part of the tensor 3 point function must vanish, or be small compared to the other two terms.}  Thus, the tensor bispectrum may be the only clear discriminant between slow-roll inflation and a general class of $SO(1,4)$ invariant models that includes HI.

The group theory analysis does not determine the scaling dimension $\Delta_{S}$ or the coefficients of the various two- and three-point functions.  In the next section, we review the Holographic Inflation (HI) model, which makes predictions for some of these unknown constants.
Note however, that Maldacena, using the bulk effective field theory description of fluctuations, has derived several relations between the nine parameters on quite general grounds.  The fundamental gauge invariant measure of scalar fluctuations is the scalar metric perturbation $\zeta$, where
\begin{eqnarray*}
ds^2 &=& - N^2 dt^2 + h_{ij} (dx^i + N^i)(dx^j + N^j)\\
h_{ij} &=& a^2(t) [(1 + 2\zeta)\delta_{ij} + \gamma_{ij} ],
\end{eqnarray*}
with $\gamma_{ij} $transverse and traceless. When $\zeta (x) $ is constant, this is just a rescaling of the spatial FRW coordinates so its effect is completely determined.  Thus, in a three-point function including $\zeta$, which depends on three momentum vectors satisfying the triangle condition ${\bf k_1 + k_2 + k_3} = 0$, the squeezed limit where the $k_i$ of $\zeta$ is taken to zero, is completely determined by the coordinate transformation of the corresponding two-point function.  Since $SO(1,4)$ fixes the momentum dependence of all $3$ point functions up to a multiplicative constant, the constants in the $\langle S T T \rangle$, $\langle SST\rangle$ and $\langle S S S \rangle$ three-point functions, are determined by those in the $\langle T T \rangle$, and $ \langle S S \rangle$ two-point functions.  This leads to the prediction of small non-Gaussianity in the slow-roll limit, and reduces our $9$ constants to $6$.  We have argued that the HST model does have a description in terms of coarse grained classical field theory, and so should obey Maldacena's constraint.

 Slow-roll inflation models determine the magnitudes of fluctuations in terms of the quantum fluctuations of canonically normalized free fields in the Bunch-Davies state.  In the single field slow-roll models, this leads to an exact relation between the scalar and tensor tilts and the normalizations of the $\langle S^2\rangle$ and $\langle T^2\rangle$ correlators.  The HI model does not lead to this relation. However, the relative orders of magnitude of the scalar and tensor two-point functions are determined by very general geometrical considerations.  The quantity $\zeta$is shown in Appendix A of\cite{lyth} to satisfy
$$\zeta = - 3 \bar{H} \delta t ,$$ where $\delta t$ is the proper time displacement between two infinitesimally separated co-moving hypersurfaces, and $\bar{H}$ the homogeneous Hubble radius.  This requires only that the metric be locally FRW, that the cosmological fluid have vanishing vorticity, and that fluctuations away from homogeneity and isotropy are treated to first order.  On the other hand, 
$$\delta t = \frac{\delta H}{\dot{\bar{H}}}, $$ where $\delta H$ is the local fluctuation in the Hubble parameter.  If the metric is close to dS then $\dot{\bar{H}}$ is small.
$\delta H$ is the fluctuation in the inverse radius of space-time scalar curvature, while the tensor fluctuations are fluctuations in the spin two part of the curvature, which is defined intrinsically by the fact that the background is spatially flat.
Thus, we can conclude quite generally that the fluctuations in $\delta H$ and in the spin two piece should be of the same order of magnitude.  In section 3 we will recall that in the HI model, general statistical arguments indicate that these fluctuations have the magnitude $\frac{1}{R M_P}$, where $R$ is the radius of the approximate dS space.  
We want to emphasize that apart from the last remark, these are purely classical geometrical considerations.  Adopting Jacobson's point of view about Einstein's equations, we can say that any quantum theory of gravity whose local hydrodynamics looks like dS space for a sufficiently long period, will give predictions for scalar and tensor fluctuations that are qualitatively similar to those of slow-roll inflation.  We will discuss the observational discrimination between different models below.

\subsection{Tilt}

The scalar two-point function is given at large times in the flat slicing by
$$ \langle \zeta (k) \zeta (-k) \rangle = \frac{A}{k^3} \frac{H^2}{\dot{H}^2} (\frac{e^{t/R}}{k})^{- 4\Delta_S} .$$ 
In slow-roll models, The relevant value of $t$, at which to evaluate this formula, depends on $k$ via the equation
$$ k = a (t(k)) H(t(k)).$$  Notice that in these formulae we've reverted to the use of $R$ for the constant inflationary dS radius, while $H$ is the varying Hubble parameter. In a general model, $H$ will be decreasing with time and $\dot{H}$ will be increasing, as inflation ends.  

Modes with higher $k$ leave the horizon at a later time and so the normalization $\frac{H^4}{\dot{H}^2}$ will be smaller for these modes.
However, there is another effect coming from the fact that $\Delta_S$ is positive.  As inflation ends, $a(t)$ is not increasing as rapidly as the exponential so $\frac{e^{t/R}}{a(t)} $ increases as $t$ increases (we neglect the variation of $H(t)$ in the horizon crossing formula, because it is not in the exponential).  Thus, the logarithmic derivative of the correlation function will have a negative contribution from the prefactor and a positive one proportional to $ \Delta_S$ .  Since both effects depend on the slow variation of $H$, the tilt will be small (remember that $ \Delta_S$ is bounded by unitarity), but its sign depends on the value of $\Delta_S$.  The slow-roll result of red tilt is obtained for small $\Delta_S$, but near the unitarity bound the tilt could be of either sign, depending on the behavior of $H(t)$.  The conventional slow-roll model usually assumes $\Delta_S =0$.

Similar remarks apply to the tensor fluctuations.  However, the overall constant in these is not in general fixed in terms of the normalization of the scalar two-point function, as it is in a conventional slow-roll model.   If we ever measure the tensor fluctuations, we will be able to see whether the slow-roll consistency condition, relating the magnitudes and tilts of tensor and scalar two-point functions is satisfied.

Thus far, we've compared slow-roll inflation to a general model satisfying only approximate $SO(1,4)$ invariance of the density matrix, and the existence of an approximately de Sitter classical background geometry $H(t)$.  If we now specialize to models constructed in HST, we find a different prediction for the scale dependence of the normalization parameter $A$ (and the corresponding normalization of the tensor two-point function).  In slow-roll models we find
$$A_{S,T} = C_{S,T} (\frac{H(t)}{M_P})^2 , $$ with fixed numerical coefficients.  The HST model, as we will explain below, predicts instead that $$A_{S,T} = D_{S,T}^{\prime} (\frac{1}{R M_P})^2 , $$ with numerical coefficients which are not yet calculable.  This has the consequence that {\it the HST model predicts no tensor tilt}.  It also suggests that the size of the tensor fluctuations might be large enough to be seen in the Planck data (but the unknown coefficients make it impossible to say this definitively).  

For a given function $H(t)$ we have the following predictions for the scalar tilt
$$ n_s^{\rm slow\ roll} = \frac{H}{H^2 + \dot{H}}\frac{d}{dt} (6 {\rm ln}\ H - 2 {\rm ln}\ (\dot{H}) ) .$$
$$ n_s^{HST} = \frac{H}{H^2 + \dot{H}}\bigl[ \frac{d}{dt} (4 {\rm ln}\ H - 2 {\rm ln}\ (\dot{H}) ) - 4\frac{\Delta_S}{R} \bigr] + 4 \Delta_S .$$
Note that in the slow-roll limit, where $H \approx R^{-1}$, the last term in square brackets cancels the term outside the brackets, and also that $\Delta_S$ is bounded from above by $3/2$.  The two formulae for the tilt
are different, but both predict that it is small, and that the sign of $n_s - 1$ depends on the time variation of $H(t)$.  Note however that $H(t)$ is not measured by anything else than the primordial fluctuations, so we can adjust $H(t)$ and $\Delta_S$ to make a slow-roll model have, within the observational errors, the same predictions as an HST model.  It's possible that further study of the consistency conditions on HST models would enable us to make more precise theoretical statements about $H(t)$ and $\Delta_S$, but at the moment it does not appear that the scalar power spectrum can distinguish between them.  The absence of tensor tilt {\it is} a clear distinguishing feature.

Measurement of the tensor bispectrum would give us a much finer discrimination between models.  {\it In particular, observation of the parity violating part of this function would rule out all models based on conventional effective field theory in the Bunch-Davies vacuum.} It's unfortunate then that HI, like the slow-roll models, predicts that the tensor bispectrum is down by a factor of $\frac{H}{m_P}$, from the tensor two-point function, which is in turn smaller than the already observed scalar fluctuations by a factor of order $\frac{\dot{H}}{H^2}$.  It seems unlikely that we will measure it in the near future.

\subsection{Comparison With the Approaches of Maldacena and McFadden-Skenderis}

Maldacena's derivation of the dS/CFT correspondence implies that the quantum theory defined by his equations carries a unitary representation of $SO(1,4)$, {\it within the semi-classical approximation to the bulk physics}.  He argues that the analytic continuation, in the c.c., of the generating functional of correlators of an Euclidean CFT with a large radius AdS dual, is the Wheeler DeWitt (WD) wave functional of the corresponding bulk Lagrangian on the 3-sphere.   This argument has been generalized to all orders in the semiclassical expansion of the bulk Lagrangian by Harlow and Stanford\cite{hs}.  {\it To all orders in the semi-classical expansion} the WD wave functional defines a positive metric Hilbert space.  The correlation functions defined by Maldacena are expectation values of operators localized at points on the sphere, in a given state in this Hilbert space.   They are covariant under $SO(1,4)$ and the Hilbert space carries a unitary representation of $SO(1,4)$.  In this semi-classical analysis the state in the Hilbert space is the Bunch Davies vacuum for dS quantum field theory, defined by analytic continuation from a Euclidean functional integral.

It is important to realize that these correlators are {\it not} correlators in the "non-unitary" CFT, which defines the coefficients in the exponent of the WD wave function.   The complex weights, which seem so mysterious in the CFT are familiar as the complex parameters labeling the complementary and principle series of {\it unitary} representations of $SO(1,4)$.  Furthermore, although our quantum theory contains operators localized at points on the 3-sphere, it is NOT a Euclidean QFT on the 3-sphere.   The correlators in such a QFT would be analytic continuations of expectation values of operators in a theory on $2 + 1$ dimensional dS space, and would satisfy reflection positivity on the 3-sphere. This cannot be the case because none of the generators of $SO(1,4)$ are bounded from below. The usual radial quantization of a theory on the sphere describes a Hilbert space composed of unitary highest weight representations of $SO(2,3)$, whose analytic continuation are highest weight non-unitary representations of $SO(1,4)$, not the unitary unbounded representations that one finds by doing bulk quantum field theory.

Many people have been tempted to use the AdS/CFT correspondence to {\it define} a quantum theory in dS space by just using the analytically continued correlators of some exact CFT to define a non-perturbative WD wave function.  We are not sure what this would mean.  The formal analytic continuation of the path integral gives rise to a wave functional satisfying the exact WD equation.  There is no positive definite scalar product on the space of solutions of this hyperbolic equation,and it is not clear how to give a quantum interpretation of the correlation functions that would be defined by this procedure.   We propose instead that the correct non-perturbative generalization of Maldacena's observation is that the inflationary correlation functions, in leading order in deviations from dS invariance, be given by expectation values of localized operators in a quantum theory on the 3-sphere, carrying a unitary representation of $SO(1,4)$.  As we've noted above, current observations are completely accounted for by this principle, without the need for a detailed model.  

The application of Maldacena's version of dS/CFT to inflation works only to leading order in the slow-roll approximation.  M-S instead begin from a correspondence between holographic RG flows and full inflationary cosmologies.  It has long been known that the equations for gravitational instantons, of which domain walls are a special case, have the form of FRW cosmologies, with positive spatial curvature, if we interpret the AdS radial direction as time.  In particular, when the Lorentzian signature potential has negative curvature in AdS space, corresponding to a Breitenlohner-Freedman allowed tachyonic direction, the cosmology asymptotes to dS space.  From the AdS point of view, such domain walls represent RG flows under perturbation of the CFT at the AdS maximum, by a relevant operator. M-S show that by performing a careful analytic continuation of fluctuations around the domain wall solution, they can write inflationary fluctuations in terms of analytically continued correlators in the QFT defined implicitly by the domain wall.  Furthermore, in another paper, they show that if the relevant operator is nearly marginal, then the analytic continuation of the formulas for correlators in the perturbed CFT, computed by conformal perturbation 
theory, produce fluctuations corresponding to a slow-roll model. That is, they obtain slow-roll correlators when these correlators are plugged into the formulae they derived in the domain wall case where the RG flow was tractable in the leading order AdS/CFT approximation.   They suggest a holographic theory of inflation, in which their formulae are applied to the correlators in a general QFT.

To leading order in slow-roll parameters, and in the bulk semi-classical approximation, the results of M-S are equivalent to those of Maldacena, though they are derived by a different method.  
Thus, we can give them a quantum mechanical interpretation, as above.
We are unsure what to say about the non-perturbative definitions of inflationary correlators, which they propose, since we do not know how to interpret them as quantum expectation values.  In Maldacena's case, the attempt to interpret the analytically continued generating function as the WD wave function, no longer produces quantum mechanics if the semi-classical approximation does not apply.  We cannot make a similarly definitive negative statement about the non-perturbative proposals of M-S, but we cannot prove that their procedure defines a quantum mechanics.  We suspect that the proposals of M-S and Maldacena are in fact equivalent to all orders in the bulk semi-classical approximation, at least for slow-roll models (Maldacena only treats slow-roll models), and that the same objections to the M-S proposal for using exact, analytically continued CFT correlators, would apply.  

We would like to opine that the term dS/CFT and the analytic continuation from AdS space are both somewhat misleading.  dS is inappropriate because we are not dealing with a theory of eternal dS space, with an entropy proportional to $R^2$.  In a stable dS space the correlators that we compute can never be measured by any local observer.  Instead, these formulae apply, approximately, to an inflationary model with a large number of e-folds of inflation. In such a model, the entropy accessible after inflation is of order $e^{3 N_e} R^2$, and these correlators are measurable by post-inflationary observers.  CFT is inappropriate in general because a CFT has a unique $SO(1,4)$ invariant density matrix.  The analytic continuation from AdS space is meaningful only in the semi-classical expansion, and in that expansion it gives the unique Bunch-Davies state of $SO(1,4)$ invariant bulk field theory. We have argued that apart from the precise slow-roll consistency relation, this does not give predictions for two-point functions that are significantly different than those provided by symmetry and general theorems alone.

A number of other authors \cite{others} have invoked ``conformal invariance" to constrain inflationary correlators.  While we agree with many of the equations proposed by these authors, we believe that we have provided the only correct interpretation of these results within the framework of an underlying quantum mechanical theory.   One interesting question that we have not resolved is the extent to which there exist ``Ward identities" relating correlators of different numbers of tensor fluctuations.  In standard slow-roll inflation, the normalization of tensor to scalar fluctuations is completely fixed (not just in order of magnitude in slow roll), by the normalization of the bulk Einstein action.  In the relationship with AdS/CFT, this normalization is ``dual" to the fact that the coefficients of the log of the WD wave function, are analytically continued correlators of the stress tensor.

In ordinary QFT, there are two ways to derive stress tensor Ward identities.  We can analytically continue relations derived from commutators and time ordered products in the Lorentzian continuation of the theory, {\it or} we can interpret stress tensor correlators as the response to variations of the metric of the Euclidean manifold on which the functional integral is defined.   We have taken pains to stress that in the proper interpretation of our $SO(1,4)$ invariant quantum theory, no analytic continuation to Lorentzian signature is allowed.  In the next section, when we review the construction of the quantum theory from HST, it will be apparent that the round metric on the 3 sphere plays a special role in the construction, and it is not clear how to define the model on a generic 3-geometry.  Consequently, we do not see how to define Ward identities beyond the semi-classical approximation to bulk geometry.

In summary, while the results of previous authors on ``dS/CFT" for the computation of inflationary correlators are correct to all orders in the bulk semi-classical expansion, they {\it do not} lead to a new non-perturbative definition of quantum gravity in an inflationary universe.  An appropriate non-perturbative generalization of these results is to assume the fluctuations may be calculated as expectation values of a scalar and tensor operator on $S^3$. The density matrix is approximately $SO(1,4)$ invariant. That is, we assume that the  quantum theory is approximately a reducible unitary representation of $SO(1,4)$.  We emphasize the word {\it approximate}
in these desiderata, because our HST model is finite dimensional, but approaches a representation of $SO(1,4)$ exponentially, as $N_e \rightarrow\infty$.  This statement should be take to refer to convergence of expectation values of a small number of operators.

The theory should also have a Jacobsonian hydrodynamic description in terms of classical fields in a space-time which is close to dS space for a long period, but allows the horizon to expand to encompass many horizon volumes of dS.  The purpose of the present paper was primarily to show that this broad framework was sufficient for understanding the observations.  Two of the authors, TB and WF, believe the HST model of \cite{holoinflation} is the only genuine model of quantum gravity that has these properties.  One need not share this belief to accept the general framework of symmetries and cosmological perturbation theory.

\section{The Holographic Inflation model}

The basic idea of HST is to formulate quantum gravity as an infinite set of independent quantum systems, with consistency relations for ``mutually accessible information". Each individual system describes the universe as seen from a given time-like world line (not always, or even usually, a geodesic), evolving in proper time along that trajectory.  The dynamics along each trajectory is constrained by causality: the evolution operator for any proper time interval factorizes as\footnote{We use notation appropriate for a Big Bang cosmology. $0$ is the time of the Big Bang. An analogous treatment of a time symmetric space-time would use an evolution operator $U(T,-T)$.} $$U(T,0) = U_{in} (T,0) \otimes U_{out} (T,0),$$ where $U_{in}$ acts only on ``the Hilbert space ${\cal H} (T, {\bf x})$ of degrees of freedom in the causal diamond determined by the past and future endpoints of the trajectory".  $U_{out} (t,0)$ operates in the tensor complement of ${\cal H} (T, {\bf x})$ in ${\cal H} (T_{max}, {\bf x})$ .   ${\bf x}$ is a label for the trajectory. The dimension of ${\cal H} (T, {\bf x})$, in the limit that it is large, determines the area of the holographic screen of the causal diamond via the Bekenstein-Hawking relation (generalized beyond black holes by Fischler, Susskind and Bousso),  $$A(T, {\bf x}) = 4 L_P^{d-2}\ {\rm ln\ dim}\ [ {\cal H} (T, {\bf x})] .$$  We will take $d = 4$ in this paper. The causal relations between different diamonds are encoded in commutation properties of operators, as in quantum field theory (QFT).  

$A(T,{\bf x})$ must not decrease as $T$ increases.  For small $T$ it will always increase.  It may reach infinity at finite $T$, as in AdS space; remain finite as $T\rightarrow\infty$, as in dS space; or asymptote to infinity with $T$.   For trajectories inside black holes, or Big Crunch universes, $T_{max}$ will be finite.  It's clear that there must be jumps in $T$, where the dimension of ${\cal H} (T, {\bf x})$, changes, and it's not likely that we need to discuss continuous interpolations between these discrete times.  In the models of this paper, the discrete jumps will be of order the Planck time.

For any time, and any pair of trajectories, we introduce a Hilbert space ${\cal O} (T, {\bf x,y})$ whose dimension encodes the information mutually accessible to detectors traveling along the two different trajectories.   ${\cal O} (T, {\bf x,y})$ is a tensor factor in both ${\cal H} (T, {\bf x})$ and ${\cal H} (T, {\bf y})$.
We define two trajectories to be nearest neighbors if 
\begin{equation*}
{\rm dim}\ {\cal O} (T, {\bf x,y}) = {\rm dim}\ {\cal H} (T - 1, {\bf x}) = {\rm dim}\ {\cal H} (T - 1, {\bf y}).
\end{equation*}
Translated into geometrical terms, this means that the space-like distance between nearest neighbor trajectories, at any time, is the Planck scale. The second equality defines what we call {\it equal area time slicing} for our cosmology.  We want the nearest neighbor relation to define a topology on the space of trajectories, which we think of as the topology of a Cauchy surface in space-time.  It is probable that it is enough to think of this space as the space of zero simplices of a $d - 1 = 3$ dimensional simplicial complex, but for ease of exposition we use a cubic lattice.  We require that ${\rm dim}\ {\cal O} (T, {\bf x,y})$ be a non-increasing function of the number of steps $d({\bf x,y})$ in the minimal lattice walk between the two-points.  

The choice of ${\rm dim}\ {\cal O} (T, {\bf x,y})$ for points which are not nearest neighbors is determined by an infinite set of dynamical consistency requirements.  Given time evolution operators and initial states in each trajectory Hilbert space, we can determine two time dependent density matrices $\rho (T, {\bf x})$ and $\rho (T, {\bf y})$ in ${\cal O} (T, {\bf x,y})$.  We require that
$$\rho (T, {\bf x}) = V(T,{\bf x,y}) \rho (T, {\bf x}) V^{\dagger} (T,{\bf x,y}),$$ with $V(T,{\bf x,y})$ unitary.  This constrains the overlap Hilbert spaces, as well as the time evolution operators and initial states.

As TB and WF have emphasized many times, the structure of space-time, both causal and conformal, is completely determined by quantum mechanics in HST, but the space-time metric is not a fluctuating quantum variable. The true variables are quantized versions of the orientation of pixels on the holographic screen.   They are sections of the spinor bundle over the screen, but in order to satisfy the Covariant Entropy Bound for a finite area screen, we restrict attention to a finite dimensional subspace of the spinor bundle, defined by an eigenvalue cutoff on the Dirac operator\cite{tbjk}.  For the geometries considered in this paper, with only four large space-time dimensions, the screen is a two sphere with radius $\sim N$ in Planck units times an internal manifold $K$ of fixed size.  The variables are a collection of $N\times N+1$ complex matrices, $\psi_i^A (P)$ one for each independent section of the cutoff spinor bundle on $K$.  Their anti-commutation relations are

$$[\psi_i^A (P), {\psi^{\dagger}}^j_B (Q)]_+ = \delta_i^j \delta^A_B Z_{PQ},$$ with appropriate commutation relations with the $Z_{PQ}$ to make this into a super-algebra with a finite dimensional unitary representation whose representation space is generated by the action of the fermionic generators.  

We will not have to use much of this formalism in the present paper, because the era of cosmic history that we are discussing is almost featureless.  The covariant entropy bound is almost saturated, with the size of deviation from its saturation related to the size of the fluctuations discussed in this paper.  We will explain this somewhat oracular remark below.

\subsection{Review of the HI Model}

We now review the model of inflation and fluctuations described in \cite{holoinflation}.
We begin with a holographic space time model of a flat FRW universe with $p=\rho$\cite{holocosmmath}, which we believe is the generic description of the early stages of any Big Bang universe.  The Big Bang hypersurface is a topological cubic lattice of observer trajectories.  The Hilbert space of {\it any} observer's causal diamond $T$ units of Planck time after the Big Bang,  has dimension ${\rm dim\ }{\cal P}^{T(T+1)}$, where ${\cal P}$ is the fundamental representation of the compactification superalgebra.  At each time the Hamiltonian is chosen from a random distribution of Hermitian matrices in this Hilbert space, with the following provisos

\begin{itemize} 
\item Every observer has the {\it same} Hamiltonian at each instant of time.
\item For large $T$, the Hamiltonian approaches\footnote{The word {\it approaches} means that the CFT can be perturbed by a random irrelevant operator.} that of a non-integrable $1 + 1$ dimensional CFT with central charge $T^2$, living on an interval of length of $o(T)$, with a cutoff of order $1/T$, in Planck units.  The bulk volume scales like $T^3$, so the bulk energy density scales like $1/T^2$, and the bulk entropy density like $1/T$, which is the Friedman equation for the $p = \rho$ FRW space-time.  The theory has no scale but the Planck scale, so the spatial curvature vanishes, and the model saturates the covariant entropy bound\cite{FSB} at all times.
\end{itemize} 

We then modify this model in the following way.  Choose two integers, $n$ and $N$ such that $1 \ll n \ll N$, which will determine the Hubble scale during inflation and the value of the Hubble scale corresponding to the observed cosmological constant, respectively.   Choose one point on the lattice to represent the origin of ``our" coordinate system.  We will treat the tilted hypercube consisting of all points a distance $\leq N/n$ lattice steps from the origin, differently than the points outside.  For these points, we stop the growth of the Hilbert space at time $n$, for a while, and allow the Hamiltonian to remain constant.  We also use $1 + 1$ dimensional conformal transformations to replace it with the same model on an interval of length $n^3$ with a cutoff of order $1/ n^3$.  In \cite{holoinflation} we argued that this was the Hamiltonian of a single horizon volume of dS space, with Hubble radius $n$.  The rescaling of the Hamiltonian should be viewed as a change of the trajectory under consideration from that of a geodesic observer in the original FRW, to that of a static observer in dS space. The Jacobsonian effective geometry corresponding to this model up to time $n$ is a $p=\rho$ FRW, which evolves to a dS space with Hubble radius $n$.  The Jacobsonian Lagrangian contains the gravitational field and a scalar field, and the dynamics of the underlying model would imply that they were both homogeneous, if we had stopped the growth of the Hilbert space everywhere in the lattice of trajectories.

Outside the tilted hypercube however, we continue to use the $p=\rho$ Hamiltonian.  In 
\cite{holoinflation}, TB and WF argued that if $n = N$ there was a consistent set of overlap rules, which had the property that points outside the hypercube were forever decoupled from those inside, in the sense that the overlaps between interior and exterior points are always empty.  The exterior Jacobsonian effective geometry corresponding to this model is a spherically symmetric black hole of radius $N$ in the $p =\rho$ geometry.  The interior geometry is not, however consistent with this unless $n = N$.  The Israel junction condition, if we insisted on a dS geometry in the interior, would require that the boundary of the hypercube be a trapped surface with Hubble radius $N$.   

We then proposed to modify the time evolution inside the hypercube to resolve this problem. Our modification is only to the Hamiltonian of a single observer at the center of the hypercube.  We do not have a fully consistent HST model, with compatible Hamiltonians for all observers, corresponding to this model.  However, since our single observer model behaves approximately like a local field theory at times $\gg n$, and QFT satisfies the HST overlap rules approximately, we expect that a full model can be constructed.  We call this single observer model, Holographic Inflation (HI).  
According to the rules of HST, the observer at the center of the HI model will be decoupled from the rest of the DOF of the universe forever.  Since there exist solutions of Einstein's equations with multiple black holes embedded in a $p =\rho$ universe, we believe that the HI model can be embedded inside a larger model, in which the central observer's finite universe eventually collides with other universes, with different values of the cosmological constant.  In \cite{holoinflation} we argued that this is one of many possible ways to solve the ``Boltzmann brain non-problem".  Since the collision time can be any time between a few times the current age of the universe, to the unimaginably long recurrence time for the first Boltzmann brain, this embedding is completely irrelevant to any observation we could conceivably make.

\begin{figure}[t!]
\centering
\includegraphics[width=.8\textwidth]{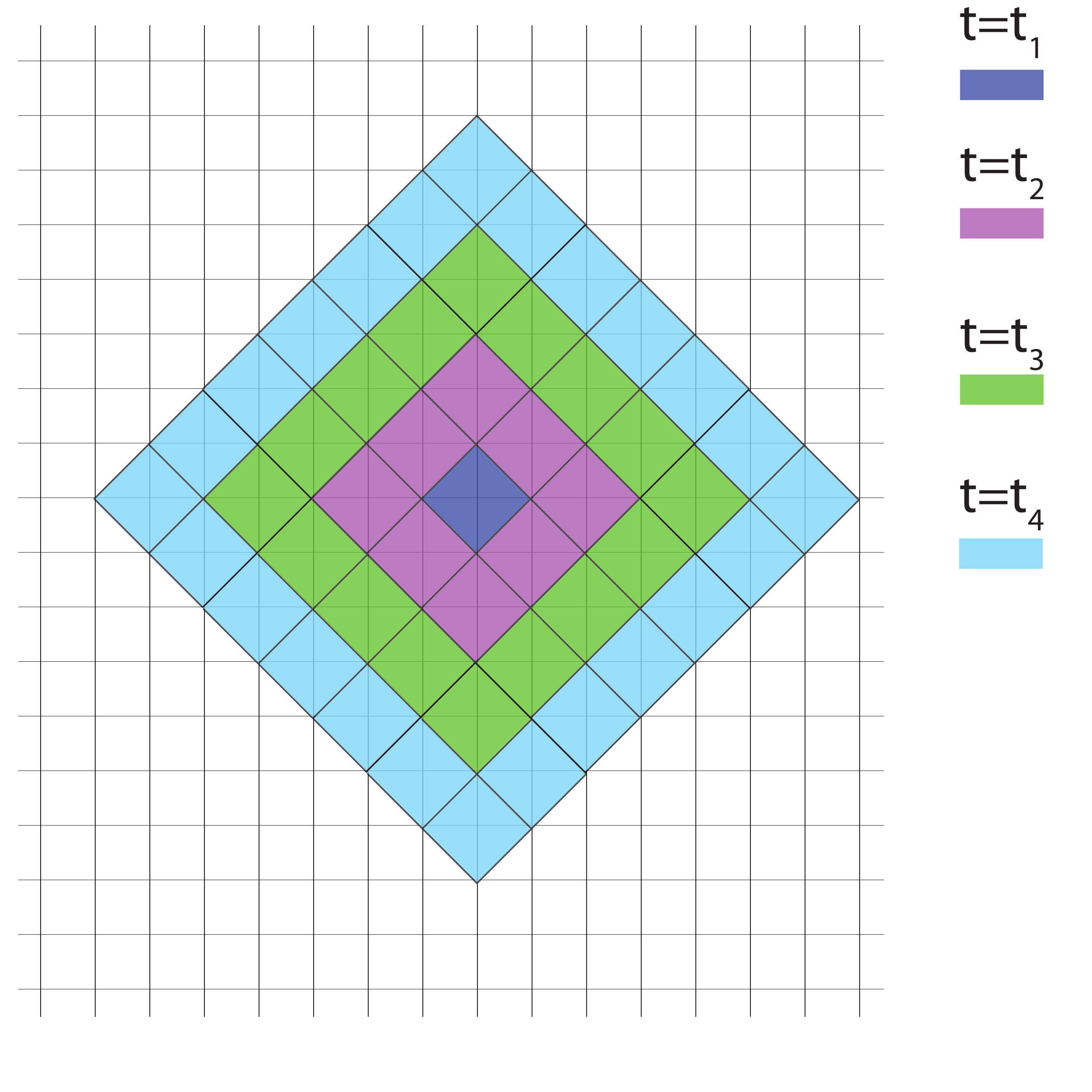}
\caption{This figure illustrates how the time dependent Hamiltonian of the HI model encompasses more DOF on the fuzzy 3-sphere (explained below), as time goes on.  Each band in the figure represents a fuzzy 2-sphere of radius $R(t_k ) = R \sin (\theta_k)$ at time $t_k$.  The horizon radius $R(t)$ is a smooth function that approximates this discrete growth of the horizon for a large number of e-folds.  It determines an FRW cosmology through
$ R(t) = R a(t)\int_{I}^t \frac{ds}{a(s)} .$.}
\label{couple}
\end{figure}

The Hilbert space of the Holographic Inflation model has entropy of order $N^2$.  Initially, the Hilbert space is broken up into $(N/n)^2 $ tensor factors, each of which
behaves like a single horizon volume of dS space.  That is to say, the state of each of these systems is changing rapidly in time in a manner that leads to scrambling of information on a time scale $n\ {\rm ln}\ n$\cite{susskindsekino}. Now we gradually begin to couple these systems together, starting from those that are close to the center of the hypercube as shown in Figure \ref{couple}.  The idea behind this is that time evolution up to time $n$ gave us multiple copies of the single $dS_n$ Hilbert space, corresponding to different observers.  We now map all of those copies into the Hilbert space of the central observer.  We want to get an emergent space-time which looks like multiple horizon volumes of $dS_n$. 

\begin{figure}[t!]
\centering
\includegraphics[width=.8\textwidth]{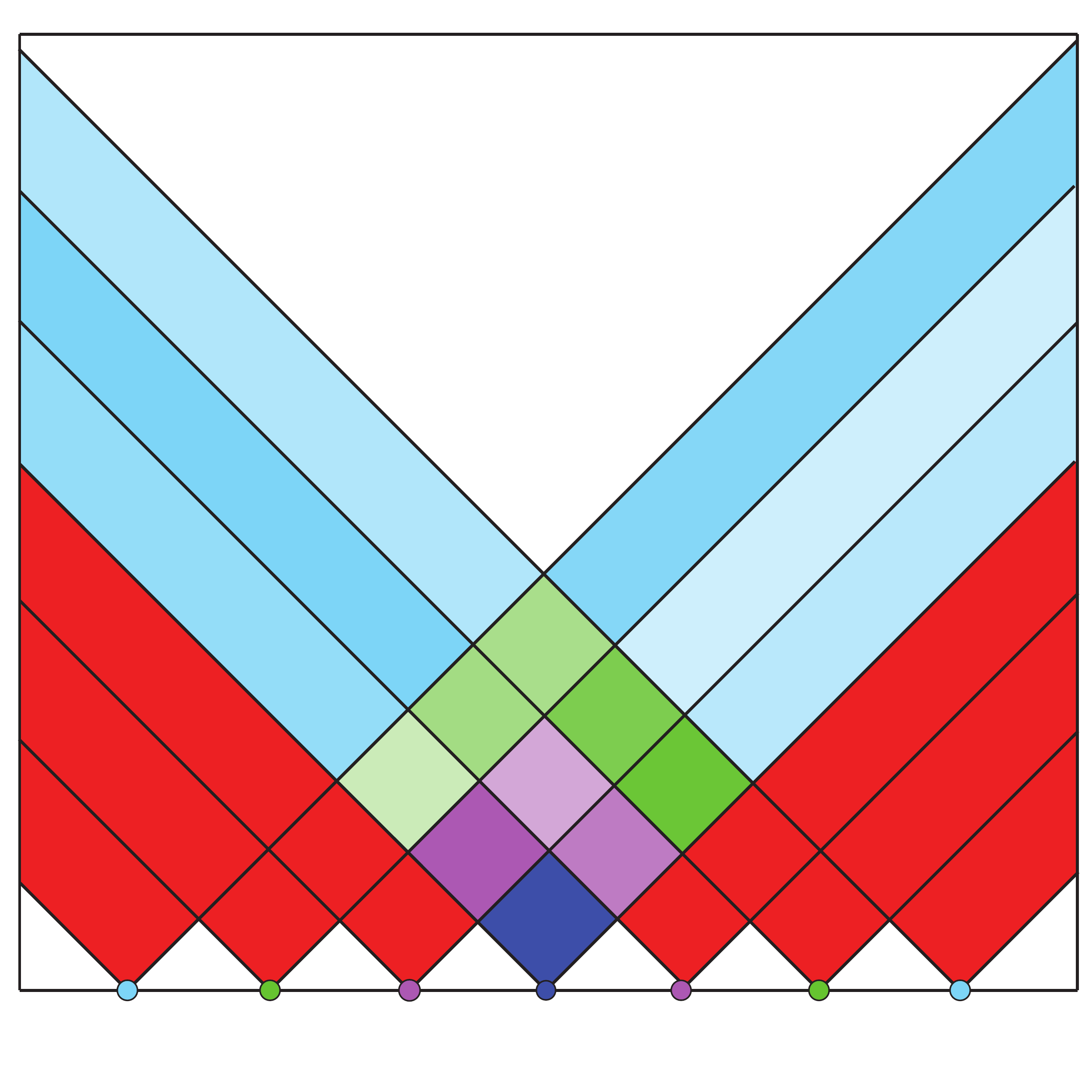}
\caption{A conformal diagram showing initially, separately evolving horizon volumes being coupled together asynchronously. The observer starts in the central horizon volume and the colored regions later in that horizon's history indicate when a nearby horizon volume (discretely separated at the colored points at $t=0$) is coupled to the Hilbert space of the observer. The red regions indicate sections of space-time that are decoupled from the central observer and allowed to evolve freely. Since this evolution is not synchronized with the time dependence of the Hamiltonian of the central horizon volume, the asynchronous coupling of independent horizon volumes gives rise to local fluctuations (indicated by different color opacities in the figure).  }
\label{conformal}
\end{figure}

Initially, the Hamiltonians of different observers were synchronized and the universe was exactly homogeneous and isotropic.  However, when we couple together the copies of these systems in the Hamiltonian of the central observer, the coupling does not occur at synchronized times.  Thus, the initial state as each successive horizon volume is coupled in can be thought of as a tensor product, but with a different, randomly chosen, state of the $dS_n$ system in each factor.  {\it This is the origin of the local fluctuations, which eventually show up in the microwave sky of the central observer.  It is also the origin of LOCALITY itself. } A conformal diagram of this unsynchronized coupling of $dS_n$ horizon volumes can be seen in Figure \ref{conformal}.

Indeed, in \cite{holoinflation} we pointed out that if we took $N = n$ then we can find a completely consistent model of a universe which evolves smoothly from the $p=\rho$ Big Bang, to $dS_N$, without ever producing a local fluctuation.  It is exactly homogeneous and isotropic at all times, despite the fact that the initial state is random and the Hamiltonian is a fast scrambler.  Although it corresponds to a coarse grained effective geometry, the model contains no local excitations around that background. Instead, it saturates the Covariant Entropy Bound at all times and is never well approximated by QUEFT, despite the fact that it is, for much of its history, a low curvature space-time.  By taking $1 \ll n \ll N$, we find a model that interpolates between the $p = \rho$ Big Bang, and asymptotic $dS_N$, via an era of small localized fluctuations, which, for a long time, remain decoupled from the majority of the horizon DOF in $dS_N$.

Thus, the role of inflation in the Holographic Inflation model is precisely to generate localized fluctuations, by starting the system off in a state where commuting copies of the same DOF are in different quantum states, from the point of view of the central observer.  Below, we will map these commuting copies to different points on a fuzzy 3-sphere, so that the fluctuations in their quantum states become local inhomogeneities of the 3-sphere.  These are, in our model, the origin of the CMB fluctuations, and they provide the raison d'{\^ etre} for localized excitations of the ultimate $dS_N$ space.  One might say that the most probable path between the $p=\rho$ geometry and $dS_N$ is the homogeneous model described in the previous paragraph.  By forcing the universe to go through a state where tensor factors of its Hilbert space are decoupled, the inflation model chooses a less probable, though more interesting, path\footnote{We are using the word probable in a somewhat peculiar way in this sentence. That is, the exactly homogeneous, entropy maximizing, model is a different choice of time dependent evolution operator than the HI model, which contains a period of inflation, and produces localized fluctuations. The latter model exploits the basic postulate of HST that the initial state in any causal diamond whose past tip is on the Big Bang hypersurface, is unentangled with DOF outside that diamond, to construct an evolution operator that exhibits approximate locality for a subset of DOF. As a consequence, the state of this model does not have maximal entropy for the period between the beginning of inflation and the time when all localized excitations decay to the dS vacuum.  It's not clear whether we should call the second model "less probable" than the first.  They are not part of the same theory.  What we mean is that, at intermediate times, a random choice of state would coincide with the actual state determined by the time dependence of the first model, while the states of the second model would look non-random.}. 

In the model described in \cite{holoinflation}, we organized all of the DOF which have interacted up to the end of inflation, in terms of variables localized on a fuzzy hemi-3-sphere of radius $E$.  In order to match with the bulk picture of inflationary geometry, this corresponds to sphere with $e^{3N_e} n^2$ DOF.  The boundary of this sphere is the holographic screen of the central observer's causal diamond at the end of inflation.
Indeed, in the bulk picture of inflation, all DOF encountered by the central observer in the future have been processed during the inflationary period.
Thus $$E^2 = e^{3N_e} n^2  \leq N^2 = 10^{123},$$ 
and $$ N_e \leq 94. 4 - 2/3 {\rm ln}\ n =   85.4 - 4/3 {\rm ln}\ \frac{M_I}{M_U} ,$$ where the ratio in the last term is that of the scale of inflation to the unification scale ($2\times 10^{16}$ GeV).

On the other hand,  $E$ must be large enough to encompass all of the degrees of freedom that manifest themselves as fluctuations in the CMB.
The entropy of CMB photons in the current universe is  
$$ (\frac{T}{M_P} N)^3 \sim 10^{89} .$$  However, the entropy in the {\it fluctuations} is only a fraction of this 
$$S_{fluct} = 3\frac{\delta T}{T} \times 10^{89} \sim e^{169} \leq e^{3N_e} \times n^2 .
$$Thus $$ N_e \geq 49 - \frac{4}{3} {\rm ln}\ \frac{M_I}{M_U}.$$ 

We can estimate the size of the local fluctuations by the usual rules of statistical mechanics.  The local subsystems have entropy of order $n^2$, so that a typical fluctuation of a local quantity, is $o(1/n)$.  This indicates an inflation scale of order the GUT scale, if we use the CMB data to normalize the two-point function. The fluctuations are also close to Gaussian, again because they are extensive on the inflationary holoscreen.  $k$-point functions scale like $n^{-k}$. Note that, apart from factors which arise from the translation of these quantum amplitudes into the fluctuations used in classical cosmological perturbation theory, this is the scaling of $k$ point functions expected in a conventional slow-roll model.  However, the size of these fluctuations is fixed by $n{-1}$, rather than by the effective $H(t)$ that one would get if one computed QUEFT fluctuations in a slowly evolving cosmology.  We pointed out in the previous section that this leads to a prediction of zero tensor tilt in the HI model, but that our ignorance of the correct form of $H(t)$ makes it difficult to differentiate the predictions for scalar tilt of the HI and bulk QUEFT models of fluctuations.

The main burden of the present paper is to explore the consequences of the dS invariance of these fluctuations.  Note that there is no meaning to dS invariance in the theory of a single stable dS space.  The physics of that system is confined to a single horizon volume, and only an $R \times SU(2)$ subgroup of $SO(1,4)$ leaves the horizon volume invariant.  The coset of this subgroup maps the observer's horizon volume into others, and does not act on physical observables.    However, in our Holographic Inflation model, the central observer sees $(E/n)^2$ horizon volumes, and if this number is large, we must build a model which closely approximates the properties of the classical $dS_n$ space, which is seen by a single observer. At the end of inflation, this observer's causal diamond contains many horizon volumes of $dS_n$ and so a model {\it approximately} invariant under $SO(1,4)$ is appropriate. We will argue that the corrections to this symmetric model, {\it for the calculation} of correlations between fluctuations at small numbers of points, are suppressed by powers of $e^{ - N_e}$, and it is reasonable to neglect them.  The continuous $SO(1,4)$ invariant model overestimates the total number of quantum states in the universe by an infinite factor, but most of those states are not probed by the limited observations we make on the CMB.

At this point it is worth noting that the model presented in \cite{holoinflation} and the present paper, does not really describe the CMB.  Within the HST formalism, we have not yet understood how to describe conventional radiation and matter dominated universes, where the source of the gravitational field is particles, rather than another effective classical field (the inflaton).  Our model ends with a time independent Hamiltonian which is (approximately) a generator, $\mathcal{L}_{04}$, of $SO(1,4)$.   It is {\it not} the Hamiltonian we have conjectured to describe particle physics in $dS_N$\cite{tbds}\cite{holounruh} . 
Thus our model is not a realistic cosmology.  Its hydrodynamic description is that of an FRW geometry coupled to a scalar field, which has small inhomogeneous fluctuations on a 3-hemisphere of radius $e^{N_e} n$.  In the Jacobsonian effective field theory description, these are fluctuations in the classical value of the inflaton, which are chosen from an approximately Gaussian, approximately dS invariant distribution described in the previous section.  The normalization of the two-point function is determined by $n$, and we've observed that it coincides with the observed strength of CMB fluctuations if $n$ is of order the unification Hubble radius, but the model has no CMB.  The Lagrangian for the inflaton must interpolate between the $p = \rho$ geometry, and $dS_N$ via a period of $N_e = \frac{2}{3} {\rm ln}\ (E/n)$ e-folds of inflation, plus sub-luminal expansion for the period when the horizon radius stretches from $E$ to $N$.  It is therefore a conventional slow-roll inflation Lagrangian, with parameters chosen to fit the underlying quantum model. 
To accommodate hypothetical HI models with blue tilt, we can either tune the inflaton potential so that the slow-roll parameter $\eta > 6\epsilon$, or use a hybrid model as the Jacobsonian THEFT.  We emphasize that from the point of view of HST, we are merely searching for the classical model that fits the hydrodynamics of an underlying quantum system.  In HI, that quantum system is not even approximately a QUEFT, at least at the beginning of inflation, when the fluctuations are actually generated. The fluctuations calculated from the density matrix of the underlying model are inserted into the classical space-time equations of the THEFT, as fluctuations of the metric, in the co-moving gauge for the inflaton.

In a more realistic model, we would have to make a transition from $\mathcal{L}_{04}$ to the Hamiltonian of a geodesic observer in $dS_N$.  The latter Hamiltonian describes particles, and we would have to show how the fluctuations in the inflaton get transmuted into distributions of photons and matter.   This is the physics encompassed in the conventional process of {\it reheating}, and the subsequent propagation of photons through an inhomogeneous space-time, including phenomena like the 
Sachs-Wolfe effect.  We know perfectly well how to build a QUEFT of this era, by coupling the classical inhomogeneous inflaton field to quantum fields describing particles.  It's basically the challenge of describing the particle physics in terms of HST that is beyond our reach at present.  There are however, a few remarks that we can make.  The first is that the conventional matter and radiation dominated eras lead to an increase in the radius of the horizon by an amount  $\alpha N$, with $\alpha$ a parameter strictly less than, but of order, $1$. The fact that $\alpha$ is less than one follows from the general properties of asymptotically dS cosmologies which are not exactly dS, while the fact that it's of order one reflects the very recent crossover between matter and radiation domination.  Thus, we should take $E \ll N$.  

\subsection{The Fuzzy 3-sphere}
In order to construct a model, which is effectively local in a 3 dimensional space, we label the $E^2 = e^{3 N_e} n^2 $  variables in the following manner.  The geometry seen by an observer at the end of inflation is a 3-sphere of radius $R_I = e^{N_e} n$.
We have of order $E^2$ degrees of freedom, which can be thought of holographically as living on the holographic screen of a causal diamond, with radius $ E \sim e^{\frac{3 N_e}{2}} n\gg R_I$, when the number of e-foldings is large.  We will distribute these ``uniformly" over a fuzzy 3-sphere of radius $R_I$. 
\begin{figure}[t!]
\centering
\includegraphics[width= .9\textwidth]{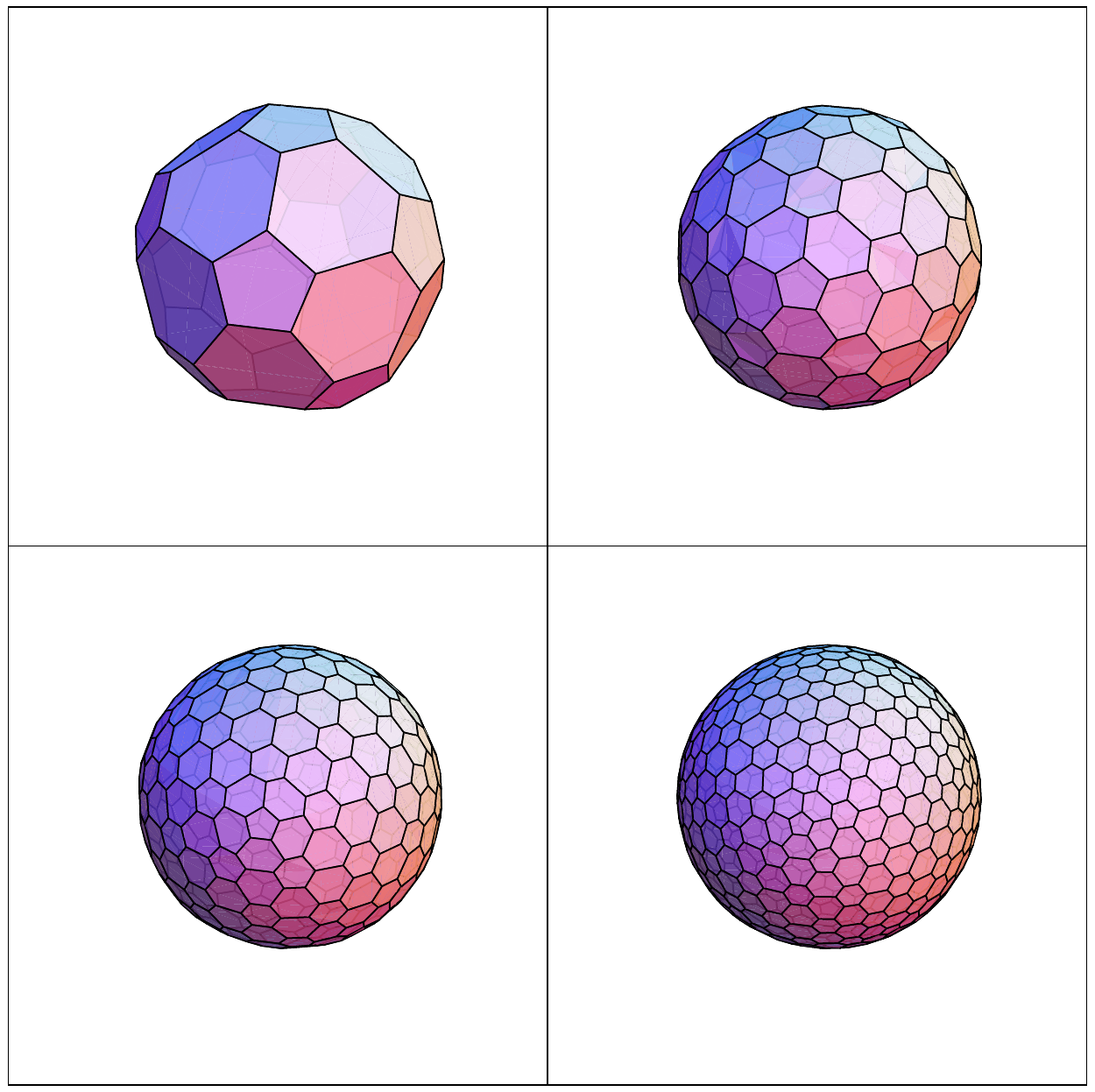}
\caption{Tilings of fuzzy two spheres of different radii.  The maximally localized spinor wave functions at the centers of the tiles are a basis for the cutoff spinor bundle, with angular momentum cutoff determined by the radius of the sphere in Planck units.}
\label{tile}
\end{figure}

 A 3-hemisphere can be thought of as a fiber bundle with two-sphere fibers, over the interval $[0, \frac{\pi}{2}]$ . The two-sphere at angle $\theta$ has a radius $R_I  \sin\theta $.   The HST version of this geometry is a collection of variables $\psi_i^A (\theta_k ) ,$ where the matrix
at $\theta_k$ is $N_k \times N_k + 1$ with $$N_k =R_I \sin\theta_k .$$
We take $\sin\theta_1 = n/R_I$, while $$\sum_k  \sin\theta_k (\sin\theta_k + 1/R_I ) = e^{N_e} .$$  The $\theta_k$ are equally spaced in angle along the interval.  Since each $N_k \geq n \gg 1$, we can construct, for each two sphere, a basis of spinor spherical harmonics localized on the faces of a truncated-icosahedral, geodesic tiling of the sphere, obtaining an approximately local description of our hemi-3-sphere. This tiling scheme is shown in Figure \ref{tile}. The centers of the faces, combined with the discretized interval parametrized by $\theta_k$ define a lattice on the 3 sphere. Our spinor variables $\psi_i^A$ have a natural action of $SO(4) = SU(2) \times SU(2)$ acting separately on the rows and columns of the matrix.  We combine this with the discrete $SO(4)$ rotations which take points of the lattice into each other.  As $N_e \rightarrow\infty$ we can construct operators which turn our unitary representation of $SO(3)$ into a unitary representation of $SO(4)$.  In addition, we argued in \cite{holoinflation} that, in the limit,  we could construct operators $\mathcal{L}_{0M}$ which extend this to a unitary representation of $SO(1,4)$.   We thus conjectured that the Hilbert space of the localized variables $\psi_i^A (\theta_k )$ admits an action of $SO(1,4)$, in the limit $N_e \rightarrow\infty$ ,
and that it can be described in terms of field operators $O_A (X)$ transforming covariantly under the action of $SO(1,4)$ on the 3-sphere, as we assumed in the previous section.  {\it We have argued that they DO NOT obey the axioms of conventional Euclidean CFT. In particular, the Hilbert space admits an infinite dimensional unitary representation of $SO(1,4)$ which cannot be highest weight (there are no highest weight unitary representations).  This also implies that there are generally many $SO(1,4)$ invariant states in the representation. Our results for the correlation functions of inflationary fluctuations depended on the assumption that the state of the system after inflation is invariant, but not on the particular choice of invariant state.}  Note that the operators $O(X)$ representing the local fluctuations commute at different points, because they probe properties of the individual, originally non-interacting, horizon volumes.  We work in the Schrodinger picture, in which the density matrix, rather than the operators, evolve. 

The preceding paragraph described mathematics.  We incorporate it into the physics of our inflationary universe in the following way.  We have followed the universe using the rules of HST from its inception until a time when the particle horizon had a size $n$.  At that time, a very large number of observers have Hilbert spaces of entropy $n^2$ and are described by identical states and Hamiltonians.  The individual Hamiltonian is that of a non-integrable, cutoff $1 + 1$ dimensional field theory whose evolution, time averaged over several e-foldings, produces a maximally uncertain density matrix.   This description extends out from a central point on the lattice of observers for a distance $N/n$, up to the surface that will eventually be our cosmological horizon.   Points on the lattice of observers that are more than $n$ steps apart, have no overlap conditions.  We make a coarser sublattice, consisting of centers of tilted cubes on the original lattice, whose Hilbert spaces have no overlaps.  We now want to describe the Hamiltonian of the central observer, from the time that individual points on the coarse sublattice thermalize, until the end of inflation.   We will {\it not} provide a complete HST description of this era, because it is currently beyond our powers.

To construct this Hamiltonian, we begin with the Hilbert space of entropy $E^2$ described above, and identify points in the coarse lattice of HST observers, with points on the fuzzy 3-sphere described above.  The central observer is identified with the point $\theta_1$ on the interval.  There is no sense in further localizing it on the fuzzy two sphere at that point, because the state in its Hilbert space is varying randomly over a Hilbert space of entropy $n^2$.  There are no localized observables at length scales smaller than $n$.  We think of this geometrically as saying that the area of the hexagon centered at this observer's position has area $n^2$.  Each point on the 3-spherical lattice has, at the beginning of this era, an identical wave function in a Hilbert space of entropy $n^2$. The time dependent Hamiltonian of the central observer now begins to couple together points on the spherical lattice, in a manner consistent with causality.  That is, as the proper time of the central observer increases, we assume that its causal diamond increases in area, and the Hamiltonian couples together points that are closest to it on the 3-sphere, in accordance with the covariant entropy bound.   In principle, the rate, in proper time, at which the area of the holographic screen grows, tells us about the FRW background geometry.  The Jacobsonian effective field theory of this is a model of gravity coupled to a scalar, with a potential that leads to $N_e$ e-folds of inflation, and a rapid transition to $dS_N$. We are dealing with only a single observer, and do not have overlap constraints to guide us, so we could incorporate any geometry consistent with the entropy bounds.  

The rate at which different points on the sphere are coupled together is not connected to the rate of change of the state according to the local Hamiltonian, which is randomizing individual Hilbert spaces of entropy $n^2$.  {\it Therefore there will be local fluctuations of the initial quantum state at different points on the 3-sphere}.  This is the physical origin of the fluctuations whose form we described in the previous section.  Above, we have argued that when $n \gg 1$ they are approximately Gaussian and estimated their magnitude.  They should clearly be thought of as statistical fluctuations in the quantum state, rather than quantum fluctuations in a pure state.  Of course, since we detect these fluctuations in properties of a macroscopic system, there is no way that one could have ever detected the quantum nature of fluctuations in the conventional inflationary picture, but the point of principle is significant.  In a more realistic model, these fluctuations would be the origin of what we observe in the CMB and the clumpy distribution of matter around us.

We construct our model so that, by the time the size of the holographic screen has reached $E$, the Hamiltonian of the DOF in that diamond is the generator $\mathcal{L}_{04}$, which approaches an element of the $SO(1,4)$ Lie Algebra in the (fictitious) limit $N_e \rightarrow\infty$.  The system is characterized by a density matrix, because the state of each point on the fuzzy 3-sphere is random, and the times at which different points become coupled together are not locked in unison\footnote{For purists, we should point out that we're not postulating non-unitary evolution, merely noting that the initial conditions of our problem introduce some randomness into the pure state of the universe.  We're simply making predictions by averaging over this ensemble of possible random states, since no observation can ever determine what the correct initial state was.}.  Note however that the initial time averaged density matrices at each point {\it are} identical, by construction, and are exactly $SO(3)$ invariant.  It is extremely plausible that the density matrix is approximately $SO(1,4)$ invariant when $N_e$ is large.   This is our principal assumption.  The ``lattice spacing" on our 3-sphere is of order $e^{- N_e}$ so corrections to $SO(1,4)$ invariance are, plausibly, exponentially small.  Note that we are free to construct a model for which this is true.  The only constraint on model building in HST (apart from those we are clearly satisfying) comes from the overlap rules.  We are not, of course, implementing the overlap rules in this paper, but we see no reason why they should be incompatible with approximate $SO(1,4)$ invariance of a single observer's density matrix.

It's important to realize that $SO(1,4)$ invariance of the density matrix does not imply exact dS invariance of the universe, as described by its THEFT. The density matrix is a probability distribution for fluctuations and the THEFT is the result of classical evolution starting from typical initial conditions.  This is, of course, exactly as in conventional inflation models.  Also, the fact that, in the underlying HI model, all degrees of freedom are in interaction, means that inflation is ending, so even the homogeneous background should be moving away from its dS form.

\section{Conclusions and Comparison With Observations}

We have argued that the form of primordial fluctuations, which has been derived to leading order in slow-roll parameters for a slow-roll inflation model with the assumption of the Bunch-Davies vacuum (see Appendix A for an argument that this assumption is a fine tuning of massive proportions), in fact follows from a much less restrictive set of assumptions.  These are $SO(1,4)$ invariance and approximate Gaussianity, plus a particular choice for the $SO(1,4)$ representations for the operator representing scalar fluctuations.  This choice, plus $8$ normalizations for the different two- and three-point functions, determine the fluctuations uniquely.  In slow-roll models, these normalizations depend on parameters in the slow-roll potential, while Gaussianity is a prediction of the model and the leading non-Gaussian amplitude is suppressed by a power of the slow-roll parameters.  We have noted that the dominance of scalar over tensor two-point fluctuations is a general consequence of cosmological perturbation theory for near de Sitter backgrounds, and the assumption that the scalar and tensor components of the curvature have similar intrinsic fluctuations (as they do in both slow-roll and HI models).  Maldacena's squeezed limit theorem, combined with $SO(1,4)$ invariance, determines all three-point functions involving scalars in terms of the scalar and tensor two-point functions.  

We've also reviewed the HST model of inflation presented in \cite{holoinflation}.  It predicts approximately Gaussian and $SO(1,4)$ invariant fluctuations, robustly and without assumptions about the initial state. Like all HST cosmologies it is completely finite and quantum mechanical.  $SO(1,4)$ invariance follows from the assumption that evolution with the $\mathcal{L}_{04}$ generator of an initially $SO(3)$ invariant density matrix will lead to an $SO(1,4)$ invariant density matrix after a large number of e-foldings.
The number of e-foldings is not a completely independent parameter, but is bounded by the ratio between the inflationary and final values of the Hubble radius.  If we require that we have enough entropy in the system at the end of inflation to account for the CMB fluctuations, then
$$ 49 - \frac{4}{3} {\rm ln}\ \frac{M_I}{M_U} \leq N_e \leq  85.4 - \frac{4}{3} {\rm ln}\ \frac{M_I}{M_U} .$$  In order to leave room for the subluminal expansion of conventional cosmology, we should not be near the lower bound.

In the slow-roll models, the small deviation from the ``scale invariant" predictions $$n_S = n_T + 1 \sim 1$$ is explained by the slow-roll condition.  A similar argument for a general $SO(1,4)$ symmetric model (and in particular the HI model) follows from the fact that the parameter $\Delta_{S}$ labeling the scalar fluctuations is bounded, $\leq 3/2$, by unitarity of the representation of $SO(1,4)$.  The construction of the HST model guarantees that the effective bulk geometry, constructed from local thermodynamics following the prescription of Jacobson, goes through a period of inflation, which ends.  We do not yet have an HST description of reheating, and the era of cosmology dominated by particle physics. The dominance of the scalar over tensor fluctuations, the smallness of non-Gaussianity involving the scalar, and the fact that the scalar and tensor tilt are both small, all follow from the fact that $\frac{\dot{\bar{H}}}{(\bar{H})^2}$ is small and that $\Delta_{S}$ is bounded.  At the level of two-point functions, the only relation that distinguishes conventional slow-roll inflation (including hybrid inflation models) from generic dS invariant quantum theory is the precise relation between the normalizations and tilts of the scalar and tensor fluctuations and the fact that the HI model predicts vanishing tensor tilt. Depending on the precise form of $H(t)$, there may be a critical value of $\Delta_S$ for which the scalar tilt shifts from red to blue.  It will be interesting to see whether further investigation of HST models can predict that the scalar tilt is red.  At our present level of understanding, the scalar tilt is a competition between a blue tilt induced by choosing a ``massive" representation for $\Delta_S$ and a red tilt induced by the conventional normalization of fluctuations.  We do not have an {\it a priori} argument for which of these dominates, or even whether there are different models where either can dominate.

At the level of non-Gaussian fluctuations, things are a bit more interesting. Slow-roll models with Lagrangians containing only the minimal number of derivative terms give rise to only one of the three possible $SO(1,4)$ covariant forms for the triple tensor correlation function.  Even if we include higher derivatives, we cannot get the parity violating form.   Thus, observation of the purely tensor bispectrum could tell us whether we were seeing conventional slow roll, or merely a generic model with approximate $SO(1,4)$ symmetry. On the other hand, the parity violating amplitude might be forbidden in general by a discrete symmetry of the HI model. At the moment, we do not see an argument, which would require such a symmetry.

We also want to emphasize that the inflation literature is replete with models which give the standard predictions for two-point functions, but predict three-point functions which are far from $SO(1,4)$ invariant.  In these models, Maldacena's squeezed limit theorem does not imply that the scalar three-point function is small everywhere in momentum space.  According to our current understanding, observation of a large scalar 3 point function, could rule out all models based on $SO(1,4)$ symmetry, and might point to some non-vanilla, QUEFT based inflation model.

Our considerations imply that so long as observations remain consistent with some slow-roll inflation model, they will not distinguish a particular model among the rather large class we have discussed without also observing tensor fluctuations.   The only observations that are likely to validate the idea of a QUEFT with Bunch-Davies fluctuations of quantum fields are a precise validation of the single field slow-roll relation between two-point functions, or a measurement of the tensor three-point function.  On the other hand, observations that validate non-standard inflationary models, like DBI inflation, or show evidence for iso-curvature fluctuations, could rule out the general framework discussed in this paper.  While it is possible that HST models can be generalized to include iso-curvature fluctuations, this is not in the spirit of those models.  The key principle of HST cosmologies is that the very early universe is in a maximally mixed state, which is constantly changing as new DOF enter the horizon.  The model of \cite{holoinflation}, was designed to be the minimal deviation from such a maximal entropy cosmology, which allowed for a period in which localized excitations decouple from the bulk of the degrees of freedom on the horizon.  A model with more structure during the inflationary era would introduce questions like ``Why was this necessary?", which could at best be justified (though it seems unlikely) by anthropic considerations.

To conclude, we want to re-iterate a few basic points.  Conventional inflation appears fine tuned because of what is usually called the Trans-Planckian problem (Appendix A).  A generic state of the DOF that QUEFT buries in the extreme UV modes of the inflationary patch, has no reason to evolve to the Bunch-Davies state.  Moreover if we accept the idea that local patches of dS space become completely thermalized within a few e-foldings, and that the generic state has no localized excitations, then it does not really make sense to treat its dynamics by QUEFT.  In \cite{holoinflation} we proposed a model that preserves causality, unitarity, the covariant entropy bound, and which, with no fine tuning of initial conditions, leads to a coarse grained space-time description as a flat FRW model with a large number of e-folds of inflation. The model produces a nearly Gaussian spectrum of almost-deSitter invariant scalar and tensor metric fluctuations.  The model can be matched to a slow-roll QUEFT model (with a different space-time metric) at the level of scalar fluctuations, but predicts no tensor tilt and, in the absence of an explicitly imposed symmetry, would have all three invariant forms of the tensor three-point function with roughly equal weights.

\appendix
\section{ Conventional inflation is fine tuned}

Inflation was originally invented to solve a number of initial condition problems in the standard theory of the Big Bang, among them the horizon and flatness problems.  The horizon problem was originally stated as ``Why is the universe homogeneous and isotropic shortly after the Big Bang, when different regions have not been in causal contact?", but also includes the question ``How can the CMB in parts of the sky with angular separation of order $\pi$ be correlated, since these represent degrees of freedom that were never in causal contact?"

The conventional inflationary answer to these questions  is that everything we see in the universe today ``originated in a tiny inflationary patch of size a few times the inflationary Hubble radius".  The way this works in QFT is that field modes localized in regions much smaller than the Planck scale, get blown up to the size of our current horizon, or larger, during the inflationary era.  This fact was emphasized in a large number of papers written in the early part of the 21st century\cite{transplanckian}, and is known as the {\it transplanckian mode problem}.  The conventional inflationary argument for dealing with these modes, whose initial description is outside the realm of validity of effective field theory, invokes the adiabatic theorem.  If those modes are in their ground state initially, then the adiabatic theorem guarantees that they will adiabatically evolve to the Bunch-Davies state when they become long wavelength enough to be treated by effective field theory.

This argument makes two crucial assumptions.  The first is the way in which the true theory of quantum gravity reduces to QUEFT.  That is, it assumes all of the mysterious modes, which show up as Transplanckian field modes in the QUEFT treatment, are indeed high energy modes of the quantum Hamiltonian, and there is a gap between their adiabatic ground state and the next excited state.   This is simply untrue in the HST model of quantum gravity, but one could assume HST is wrong, and that the real theory of quantum gravity will obey the adiabatic argument.  

The second assumption is however an admission of fine tuning worse than those that inflation was supposed to solve.  Any model of inflation that purports to explain the correlations in the CMB fluctuations, must have a Hilbert space of dimension at least $e^{(10)^{77}}$.  The dimension of the space of field theory states in a single horizon volume, which do not back-react on the metric and cause the creation of a horizon filling black hole is at most
$e^{k n^{3/2}}$, where $k$ is a constant of order $1$ and $n$ is the inflationary Hubble scale in Planck units.  The vast bulk of the Hilbert space required to explain the CMB cannot be modeled by field theory in a single $dS_n$ horizon volume.  To claim that the initial state is ``the unique adiabatic ground state" is a ridiculous fine tuning, even if the Hamiltonian of quantum gravity had the gap that is assumed in the adiabatic argument.  Generic states of large quantum systems lie in a quasi continuum of closely spaced chaotic levels.  This is the property that makes statistical mechanics work.  Adiabatic arguments for the detailed quantum state are never valid, though there are often collective variables which do evolve adiabatically. The collective variables are typically classical because they're averages over many degrees of freedom.  

The basic problem here is that the naive inflation idea that everything came from a few inflationary Hubble volumes, an inflationary Hubble time after the initial singularity, violates the Covariant Entropy Bound.  The causal diamonds of observers on that time-slice are too small to encompass the states of the observable CMB.  Field theory performs the miracle of inflation by sucking up degrees of freedom from the extreme UV, but no theory of quantum gravity, which obeys the CEB, can do the same.  What we're arguing in this appendix is that there is no magic universality argument by which inflation can capture the unknown dynamics of the non-field theoretic DOF and reach the conclusion that the Bunch Davies state is the state the universe is drawn to.  

Note that there is no contradiction here with the claim that the B-D vacuum is a good approximate description of eternal dS space.  If one accepts the contention of Banks and Fischler that dS space is a quantum system with a finite number of states\cite{tbwf}, then many of the properties of BD follow from the principles of statistical mechanics\cite{tbds}, and some mild assumptions about the Hamiltonian. 

TB has presented this argument about fine tuning in conventional inflation at a number of conferences since 2007, and it has been greeted with a great deal of skepticism.  Much of that had to do with the belief that the conventional theory of inflation had been verified by observation of the CMB.  In this paper we've shown that, to a large extent, those predictions follow from symmetries and the rather natural assumption that fluctuations are small and approximately Gaussian.  The Holographic Inflation model in fact predicts both of those two facts in a simple manner, and does not suffer from any fine tuning of initial conditions.

An independent argument that it is invalid to treat the inflationary era in the QUEFT approximation, comes from a somewhat more holographic perspective.  Everyone would agree that after a few e-folds of inflation, any given horizon volume of the universe looks like empty de Sitter space.  This is indeed the most naive argument, within QUEFT, for assuming the Bunch-Davies state.  However, a horizon volume of empty dS space is a highly entropic system, with an entropy that cannot be understood in terms of localized excitations in the bulk of the horizon volume.  If we indeed want our model of inflation to resemble a collection of decoupled horizon volumes of dS space, then we should not model its fluctuations on those of localized degrees of freedom in a horizon volume.  The HST model of inflation indeed models the local DOF as a generic thermalized system, with the right entropy and no intra-horizon locality.  Locality on super-dS-horizon scales comes from our explicit construction of a Hamiltonian based on the geometry of a fuzzy 3-sphere, to couple together the individual horizon volumes.  We had originally imagined some sort of duality between the HST description and the conventional QUEFT description, with a UV cutoff at the dS-horizon scale.  However, when we realized that most of the predictions could be understood without QUEFT, and without the assumption of the Bunch-Davies state, and that our detailed predictions for the magnitude of fluctuations differed from those of QUEFT, we abandoned this idea.  Our current thinking is that the standard approach reproduces the data, only because it obeys the symmetry principles described in this paper.

\section{Unitary Representations of $SO(1,4)$ }
Here we wish to remind the reader of the classification\cite{thomasnewton} of the unitary representations of $SO(1,4)$ by labeling representations of the group according to their eigenvalues of the invariant Casimir operators. To start, we review by constructing elements of the algebra $\mathfrak{so}(1,4)$ explicitly, wherein we can separate the elements into a 6-dimensional sub algebra of rotations in $\mathbb{R}^4$ and 4 ``boosts," which we interpret as rotations in the plane of one of the directions  $\{w,x,y,z \} \in\mathbb{R}^4$ and the ``time" coordinate, $t$, in 5d Minkowski space. The elements of the algebra are then simply  the generators of motion for the isometries of the 4d hyper-surfaces defined by the equation $t^2-(w^2+x^2+y^2+z^2)= R^2$ in 5d Minkowski space. These hyper-surfaces  define hyperboloids of constant positive curvature embedded in the underlying Minkowski space, which is of course DeSitter space. 

The elements of $\mathfrak{so}(4,1)$ are representable as linear differential operators and labeled by two greek indices that describe the plane of rotation. Thus they are 10 antisymmetric two-tensors $\mathcal{L}_{\alpha \beta}$ with $\alpha,\beta \in \{0,..,4\} \Rightarrow \{t,x,y,z,w\}$:

\begin{eqnarray*}
\mathcal{L}_{\alpha \beta} &=& \mathcal{M}_{\alpha \beta} + \mathcal{S}_{\alpha \beta}\\
\end{eqnarray*}
where the $\mathcal{M}_{\alpha \beta}$ and $\mathcal{S}_{\alpha \beta}$ represent the orbital piece and the spinorial piece respectively. 
Here the orbital and spinorial generators are constructed explicitly via 
\begin{eqnarray*}
\mathcal{M}_{\alpha \beta} &=& i ( x_\alpha \partial_\beta - x_\beta \partial_\alpha )\\
\mathcal{S}_{\alpha \beta} &=& \frac{i}{4} \left[ \Gamma_{\alpha}, \Gamma_{\beta} \right]\\
\left\{ \Gamma_{\alpha}, \Gamma_{\beta} \right\} &=&  2 \eta_{\alpha \beta} \mathbb{I}
\end{eqnarray*}
where the $\Gamma_{\alpha}$ are the $SO(1,4)$ gamma matrices, and thus the $\mathcal{L}_{\alpha \beta}$ obey the algebra
\begin{equation*}
[\mathcal{L}_{\alpha \beta} , \mathcal{L}_{\mu \nu}] =  i ( \eta_{\beta \mu} \mathcal{L}_{\alpha \nu} - \eta_{\alpha \mu} \mathcal{L}_{\beta \nu} + \eta_{\alpha \nu} \mathcal{L}_{\beta \mu} - \eta_{\beta \nu} \mathcal{L}_{\alpha \mu})
\end{equation*}
Moreover, the spin piece acts on a generic rank-n tensor field as 
\begin{equation*}
\mathcal{S}_{\alpha \beta} \mathcal{T}_{\mu_1 ... \mu_n} = i \sum_{i=1}^{n} \left[ \eta_{\alpha \mu_i} \mathcal{T}_{\mu_1 ...\mu_{i-1} \beta ...  \mu_n} -\eta_{\beta \mu_i} \mathcal{T}_{\mu_1 ...\mu_{i-1} \alpha ...  \mu_n} \right] .
\end{equation*}

In five dimensions there are two possible invariants that comprise the center of the algebra and they are:
\begin{eqnarray*}
Q &=& -\frac{1}{2} \mathcal{M}_{\mu \nu} \mathcal{M}^{\mu \nu} \\
W &=& \frac{1}{8} \epsilon^{\alpha \beta \mu \nu \gamma} \epsilon_{\alpha \rho \sigma \delta \eta} \mathcal{M}_{\beta \mu} \mathcal{M}_{\nu \gamma} \mathcal{M}^{\rho \sigma} \mathcal{M}^{\delta \eta} 
\end{eqnarray*}

To classify and label the unitary representations one proceeds as follows:
\begin{enumerate}
\item Take the sub-algebra of generators of rotations in $\mathbb{R}^4$. This is an $\mathfrak{so}(4)$ sub-algebra. And since $\mathfrak{so}(4) \cong \mathfrak{su}(2) \otimes   \mathfrak{su}(2)$, then we can split the rotation sub-algebra into two unitary sub-algebras. After doing so, we have a simple method of labeling our finite dimensional representations via angular momentum eigenvalues for the two separate $\mathfrak{su}(2)$ sub-algebras. We label these angular momentum eigenvalues with $j_1$ and $j_2$ for each respective Casimir invariant $J_{(i)} = \sqrt{(J_{(i)}^1)^2 +  (J_{(i)}^2)^2 + (J_{(i)}^3)^2} $ where here the $(i)$ labels the unitary sub-algebras. 
\item Since the algebra is non-compact, we know that the unitary representations will be infinite dimensional. Following Thomas, we can come to a full labeling scheme by decomposing the Hilbert space such that it is an infinite direct sum of finite dimensional irreducible representations of the $\mathfrak{so}(4)$ sub-algebra.  Then, we can arrive at a set of linear equations relating matrix valued functions of $j_1$ and $j_2$ by examining the action of the ``boosts" via their commutation relations with the generators of the $\mathfrak{su}(2)$ sub-algebras.

\item Finally, we solve a set of recurrence relations to find the general forms of the matrix valued functions of $j_1$ and $j_2$. Then we insist that our representations be unitary, by insisting that operators in the Hilbert space be Hermitian. This gives additional constraints on the matrix valued functions such that (when related to the Casimir invariants of $SO(1,4)$) we may distinguish the unitary representations by the eigenvalues of the Casimir invariants $Q$ and $W$.

\end{enumerate}
The systematics of the above is actually rather tedious, we quote the results of the analysis as codified by Newton\\
\\
There are four classes of unitary representations:
\subsection{Classes}
\subsubsection{Class I}
\begin{eqnarray*}
Q &>& 0 \\
W &=& 0
\end{eqnarray*}
\begin{equation*}
\left. \begin{aligned}
	j_1 + j_2 =  0 \\
	j_1 - j_2 = 0
	\end{aligned}
\right \} , 
\left. \begin{aligned}
	1 \\
	0
	\end{aligned}
\right \} , 
\left. \begin{aligned}
	2 \\
	0
	\end{aligned}
\right \} , \cdots
\end{equation*}
\subsubsection{Class II}
\begin{eqnarray*}
Q &=& -(n-1)(n+2) \qquad n=1,2,3, \cdots \\
W &=& 0
\end{eqnarray*}
\begin{equation*}
\left. \begin{aligned}
	j_1 + j_2 =  n \\
	j_1 - j_2 = 0
	\end{aligned}
\right \} , 
\left. \begin{aligned}
	n+1 \\
	0
	\end{aligned}
\right \} , 
\left. \begin{aligned}
	n+2 \\
	0
	\end{aligned}
\right \} , \cdots
\end{equation*}
Where there is one possible representations for each value of $n$. 
\subsubsection{Class III}
\begin{eqnarray*}
Q &\geq& 1- \left( s+ \frac{1}{2} \right)^2 \qquad s=\frac{1}{2},1,\frac{3}{2}, \cdots \\
W &=& s(s+1)Q +(s-1)s(s+1)(s+2)
\end{eqnarray*}
Where there is one representation for each Q and s assigned, and Q can take any real value that is above the bound. 
\begin{equation*}
\left. \begin{aligned}
	j_1 + j_2 &=&  s \\
	j_1 - j_2 &=&  -s, -s+1, \cdots , s
	\end{aligned}
\right \} , 
\left. \begin{aligned}
	s+1 \\
	-s, -s+1, \cdots , s
	\end{aligned}
\right \} , \cdots
\end{equation*}
\subsubsection{Class IVa}
\begin{eqnarray*}
Q &=& -t(t+1) -(s-1)(s+2) \qquad s=\frac{3}{2},2,\frac{5}{2}, \cdots \\
W &=& -t(t+1)s(s+1) \qquad 0< t \leq s-1 
\end{eqnarray*}
With $t$ and integer or half odd integer, following the choice made for $s$.
\begin{equation*}
\left. \begin{aligned}
	j_1 + j_2 &=&  s \\
	j_1 - j_2 &=&  -s, -s+1, \cdots , -t
	\end{aligned}
\right \} , 
\left. \begin{aligned}
	s+1 \\
	-s, -s+1, \cdots , -t
	\end{aligned}
\right \} , \cdots
\end{equation*}
\subsubsection{Class IVb}
\begin{eqnarray*}
Q &=& -t(t+1) -(s-1)(s+2) \qquad s=\frac{3}{2},2,\frac{5}{2}, \cdots \\
W &=& -t(t+1)s(s+1) \qquad 0< t \leq s-1 
\end{eqnarray*}

\begin{equation*}
\left. \begin{aligned}
	j_1 + j_2 &=&  s \\
	j_1 - j_2 &=&  t, t+1, \cdots , s
	\end{aligned}
\right \} , 
\left. \begin{aligned}
	s+1 \\
	t, t+1, \cdots , s
	\end{aligned}
\right \} , \cdots
\end{equation*}
With $t$ and integer or half odd integer, following the choice made for $s$.
\subsection{Scalar and Tensor Classification}

The only class that would admit a spin zero (scalar) fluctuation would be Class I, where the mass is unrestricted so long as it is greater than zero.  The zero mass case is subtle, because the minimally coupled massless scalar in dS space has infrared issues.  However, it does give rise to a well defined Wheeler-DeWitt wave function, and the resulting correlation functions are just the $\Delta_S =0$ limit of those for massive scalars.  In any event, data will never be able to prove that $\Delta_S$ is exactly zero, and the whole framework of $SO(1,4)$ invariance for inflation is only an approximation based on the formal limit of an infinite number of e-foldings.  

For Weyl Curvature fluctuations, the only representations that are allowed are the class $IV_{a,b}$ representations with $s=2$ \cite{SO14}.  Nothing else gives rise to a transverse symmetric trace-less two tensor in the flat coordinates.
The two different class $IV$ representations are the two helicity modes of the gravitational wave field. For the scalar case we've seen in the text that a formulation on the five dimensional projective light cone provides the simplest derivation of the correlation functions.  We hope to explore an analogous formulation for the tensor fluctuations in a future publication.



\begin{thebibliography}{99}
\bibitem{holoinflation}  T.~Banks and W.~Fischler,
  ``Holographic Theories of Inflation and Fluctuations,''
  arXiv:1111.4948 [hep-th].


\bibitem{malda}  J.~M.~Maldacena and G.~L.~Pimentel,
  ``On graviton non-Gaussianities during inflation,''
  JHEP {\bf 1109}, 045 (2011)
  [arXiv:1104.2846 [hep-th]].
  ;\\  J.~M.~Maldacena,
  ``Non-Gaussian features of primordial fluctuations in single field inflationary models,''
  JHEP {\bf 0305}, 013 (2003)
  [astro-ph/0210603].
  \bibitem{mcfaddenskenderis} A.~Bzowski, P.~McFadden and K.~Skenderis,
  ``Holography for inflation using conformal perturbation theory,''
  arXiv:1211.4550 [hep-th].
 ;\\ A.~Bzowski, P.~McFadden and K.~Skenderis,
  ``Holographic predictions for cosmological 3-point functions,''
  JHEP {\bf 1203}, 091 (2012)
  [arXiv:1112.1967 [hep-th]].
  ;\\ A.~Bzowski, P.~McFadden and K.~Skenderis,
  ``Holography for inflation using conformal perturbation theory,''
  arXiv:1211.4550 [hep-th].
  ;\\  R.~Easther, R.~Flauger, P.~McFadden and K.~Skenderis,
  ``Constraining holographic inflation with WMAP,''
  JCAP {\bf 1109}, 030 (2011)
  [arXiv:1104.2040 [astro-ph.CO]].
  ;\\  P.~McFadden and K.~Skenderis,
  ``Holographic Non-Gaussianity,''
  JCAP {\bf 1105}, 013 (2011)
  [arXiv:1011.0452 [hep-th]].
  ;\\  P.~McFadden and K.~Skenderis,
  ``Observational signatures of holographic models of inflation,''
  arXiv:1010.0244 [hep-th].
  ;\\  P.~McFadden and K.~Skenderis,
  ``The Holographic Universe,''
  J.\ Phys.\ Conf.\ Ser.\  {\bf 222}, 012007 (2010)
  [arXiv:1001.2007 [hep-th]].
  ;\\ P.~McFadden and K.~Skenderis,
  Phys.\ Rev.\ D {\bf 81}, 021301 (2010)
  [arXiv:0907.5542 [hep-th]].
\bibitem{holocosm}  T.~Banks and W.~Fischler,
  ``The holographic approach to cosmology,''
  hep-th/0412097;
 \\ T.~Banks, W.~Fischler and L.~Mannelli,
  ``Microscopic quantum mechanics of the p = rho universe,''
  Phys.\ Rev.\ D {\bf 71}, 123514 (2005)
  [hep-th/0408076]
  ;\\  T.~Banks and W.~Fischler,
  ``Holographic cosmology,''
  hep-th/0405200.
  ;\\  T.~Banks and W.~Fischler,
  ``Holographic cosmology 3.0,''
  Phys.\ Scripta T {\bf 117}, 56 (2005)
  [hep-th/0310288].
  ;\\ T.~Banks and W.~Fischler,
  ``An Upper bound on the number of e-foldings,''
  astro-ph/0307459.
  ;\\ T.~Banks and W.~Fischler,
  ``An Holographic cosmology,''
  hep-th/0111142.
  ;\\  T.~Banks and W.~Fischler,
  ``M theory observables for cosmological space-times,''
  hep-th/0102077.
\bibitem{ted}  T.~Jacobson,
  ``Thermodynamics of space-time: The Einstein equation of state,''
  Phys.\ Rev.\ Lett.\  {\bf 75}, 1260 (1995)
  [gr-qc/9504004].

\bibitem{others}  I.~Mata, S.~Raju and S.~Trivedi,
  ``CMB from CFT,''
  arXiv:1211.5482 [hep-th];
  \\  J.~Garriga and Y.~Urakawa,
  ``Inflation and deformation of conformal field theory,''
  arXiv:1303.5997 [hep-th].

\bibitem{thomasnewton}
 L.~H.~Thomas, ``On Unitary Represntations of the Group of De Sitter Space,"
 Ann.\ Math. {\bf 42} 113--126 (1941);\\
 T.~D.~Newton, ``A Note on the Representations of the De Sitter Group,"
 Ann.\ Math. {\bf 51} 730--733 (1950);\\
 Jacob~G.~Kuriyan, N.~Mukunda, and E.~C.~G.~Sudarshan, 
 "Master Analytic Representations and Unified Representation Theory of Certain Orthogonal and Pseudo-Orthogonal Groups,"
 Commun.\ Math.\ Phys. {\bf 8}, 204-227 (1968).
 
 \bibitem{SO14}
 T.~Garidi, J-P.~Gazeau, and M.~V.~Takook, ```Massive' spin-2 field in de Sitter space"
J.\ Math.\ Phys. {\bf 44}, 3838 (2003), DOI:10.1063/1.1599055

 
\bibitem{tbjk}  T.~Banks and J.~Kehayias,
  ``Fuzzy Geometry via the Spinor Bundle, with Applications to Holographic Space-time and Matrix Theory,''
  Phys.\ Rev.\ D {\bf 84}, 086008 (2011)
  [arXiv:1106.1179 [hep-th]].
\bibitem{parityviolating} J.~Soda, H.~Kodama and M.~Nozawa,
  ``Parity Violation in Graviton Non-gaussianity,''
  JHEP {\bf 1108}, 067 (2011)
  [arXiv:1106.3228 [hep-th]];
  \\  M.~Shiraishi, D.~Nitta and S.~Yokoyama,
  ``Parity Violation of Gravitons in the CMB Bispectrum,''
  Prog.\ Theor.\ Phys.\  {\bf 126}, 937 (2011)
  [arXiv:1108.0175 [astro-ph.CO]].
\bibitem{lyth}   A.~R.~Liddle and D.~H.~Lyth,
  ``The Cold dark matter density perturbation,''
  Phys.\ Rept.\  {\bf 231}, 1 (1993)
  [astro-ph/9303019].

\bibitem{hs} D.~Harlow and D.~Stanford,
  ``Operator Dictionaries and Wave Functions in AdS/CFT and dS/CFT,''
  arXiv:1104.2621 [hep-th].
\bibitem{holocosmmath}  T.~Banks, W.~Fischler and L.~Mannelli,
  ``Microscopic quantum mechanics of the p = rho universe,''
  Phys.\ Rev.\ D {\bf 71}, 123514 (2005)
  [hep-th/0408076].

  
\bibitem{FSB} 

 W.~Fischler and L.~Susskind,
  ``Holography and cosmology,''
  hep-th/9806039;
  \\ R.~Bousso,
  ``The Holographic principle,''
  Rev.\ Mod.\ Phys.\  {\bf 74}, 825 (2002)
  [hep-th/0203101], and references therein.
 \bibitem{susskindsekino}  
 Y.~Sekino and L.~Susskind,
  ``Fast Scramblers,''
  JHEP {\bf 0810}, 065 (2008)
  [arXiv:0808.2096 [hep-th]].
  \bibitem{tbwf}  W.~Fischler, ``Taking de Sitter seriously". Talk given at "Role of Scaling Laws in Physics and Biology (Celebrating
the 60th Birthday of Geoffrey West"), Santa Fe, Dec. 2000. ;\\
T.~Banks,
  ``Cosmological breaking of supersymmetry? or Little lambda goes back to the future 2,''
  hep-th/0007146;\\
   QuantuMechanics and CosMology, Talk given at the festschrift for L.Susskind, Stanford University, May 2000; T. Banks, arXiv:hep- th/0007146.

  
\bibitem{tbds}  T.~Banks,
  ``Holographic Space-Time: The Takeaway,''
  arXiv:1109.2435 [hep-th];
  \\ T.~Banks,
  ``TASI Lectures on Holographic Space-Time, SUSY and Gravitational Effective Field Theory,''
  arXiv:1007.4001 [hep-th];
  \\  T.~Banks and J.~-F.~Fortin,
  ``Tunneling Constraints on Effective Theories of Stable de Sitter Space,''
  Phys.\ Rev.\ D {\bf 80}, 075002 (2009)
  [arXiv:0906.3714 [hep-th]];
  \\  T.~Banks,
  ``Holographic Space-time from the Big Bang to the de Sitter era,''
  J.\ Phys.\ A {\bf 42}, 304002 (2009)
  [arXiv:0809.3951 [hep-th]];
  \\  T.~Banks, B.~Fiol and A.~Morisse,
  ``Towards a quantum theory of de Sitter space,''
  JHEP {\bf 0612}, 004 (2006)
  [hep-th/0609062];
  \\ T.~Banks,
  ``Some thoughts on the quantum theory of stable de Sitter space,''
  hep-th/0503066;
  \\  T.~Banks,
  ``SUSY and the holographic screens,''
  hep-th/0305163.
  \bibitem{holounruh} T.~Banks and W.~Fischler,
  ``Holographic Theory of Accelerated Observers, the S-matrix, and the Emergence of Effective Field Theory,''
  arXiv:1301.5924 [hep-th].
\bibitem{transplanckian} R.~Easther, B.~R.~Greene, W.~H.~Kinney and G.~Shiu,
  ``A Generic estimate of transPlanckian modifications to the primordial power spectrum in inflation,''
  Phys.\ Rev.\ D {\bf 66}, 023518 (2002)
  [hep-th/0204129];
  \\ J.~Martin and R.~H.~Brandenberger,
  ``The TransPlanckian problem of inflationary cosmology,''
  Phys.\ Rev.\ D {\bf 63}, 123501 (2001)
  [hep-th/0005209];
  \\ J.~Martin and R.~H.~Brandenberger,
  ``A Cosmological window on transPlanckian physics,''
  astro-ph/0012031;
  \\ T.~Tanaka,
  ``A Comment on transPlanckian physics in inflationary universe,''
  astro-ph/0012431;
  \\  J.~C.~Niemeyer and R.~Parentani,
  ``Transplanckian dispersion and scale invariance of inflationary perturbations,''
  Phys.\ Rev.\ D {\bf 64}, 101301 (2001)
  [astro-ph/0101451];
  \\ A.~A.~Starobinsky,
  ``Robustness of the inflationary perturbation spectrum to transPlanckian physics,''
  Pisma Zh.\ Eksp.\ Teor.\ Fiz.\  {\bf 73}, 415 (2001)
  [JETP Lett.\  {\bf 73}, 371 (2001)]
  [astro-ph/0104043];
  \\ M.~Lemoine, M.~Lubo, J.~Martin and J.~-P.~Uzan,
  ``The Stress energy tensor for transPlanckian cosmology,''
  Phys.\ Rev.\ D {\bf 65}, 023510 (2002)
  [hep-th/0109128];
  \\  J.~Martin and R.~H.~Brandenberger,
  ``The Corley-Jacobson dispersion relation and transPlanckian inflation,''
  Phys.\ Rev.\ D {\bf 65}, 103514 (2002)
  [hep-th/0201189];
  \\  U.~H.~Danielsson,
  ``A Note on inflation and transPlanckian physics,''
  Phys.\ Rev.\ D {\bf 66}, 023511 (2002)
  [hep-th/0203198];
  \\  S.~F.~Hassan and M.~S.~Sloth,
  ``TransPlanckian effects in inflationary cosmology and the modified uncertainty principle,''
  Nucl.\ Phys.\ B {\bf 674}, 434 (2003)
  [hep-th/0204110];
  \\  R.~Easther, W.~HKinney and H.~Peiris,
  ``Observing trans-Planckian signatures in the cosmic microwave background,''
  JCAP {\bf 0505}, 009 (2005)
  [astro-ph/0412613];
  \\ K.~Schalm, G.~Shiu and J.~P.~van der Schaar,
  ``The Cosmological vacuum ambiguity, effective actions, and transplanckian effects in inflation,''
  AIP Conf.\ Proc.\  {\bf 743}, 362 (2005)
  [hep-th/0412288];
  \\  H.~Collins and R.~Holman,
  ``Renormalization of initial conditions and the trans-Planckian problem of inflation,''
  Phys.\ Rev.\ D {\bf 71}, 085009 (2005)
  [hep-th/0501158].
 \end{thebibliography}


\end{document}